\documentclass{aa_old}
\usepackage[varg]{txfonts}
\usepackage[utf8]{inputenc}
\usepackage{wasysym}
\usepackage{natbib}
\usepackage{graphicx}
\usepackage{gensymb}
\usepackage{textcomp}
\usepackage{hyperref}
\usepackage{booktabs}
\usepackage{enumitem}
\usepackage{amsmath}
\usepackage{xcolor}
\usepackage{placeins}
\usepackage{stfloats}

\bibpunct{(}{)}{;}{a}{}{,}

\begin{document}

\defcitealias{rautio2022}{Paper~I}

\title{\textsc{Cloudy} modeling suggests a diversity of ionization mechanisms for diffuse extraplanar gas}

\author{R.~P.~V.~Rautio \inst{\ref{inst1}}
  \and H.~Salo \inst{\ref{inst1}}
  \and A.~E.~Watkins \inst{\ref{inst2}}
  \and S.~Comer\'on \inst{\ref{inst3},\ref{inst4}}
  \and A.~Venhola \inst{\ref{inst1}}}

\institute{Space Physics and Astronomy research unit, University of Oulu, 90014 Oulu, Finland \\\email{riku.rautio93@gmail.com} \label{inst1}
  \and Centre of Astrophysics Research, School of Physics, Astronomy and Mathematics, University of Hertfordshire, Hatfield AL10 9AB, UK \label{inst2}
  \and Departamento de Astrof\'isica, Universidad de La Laguna, 38200 La Laguna, Tenerife, Spain \label{inst3}
\and Instituto de Astrof\'isica de Canarias, 38205 La Laguna, Tenerife, Spain \label{inst4}}

\abstract{The ionization of diffuse gas located far above the energetic midplane OB stars poses a challenge to the commonly accepted notion that radiation from OB stars is the primary ionization source for gas in galaxies.}
         {We investigated the sources of ionizing radiation, specifically leaking midplane H\,\textsc{ii} regions and/or in situ hot low-mass evolved stars (HOLMES), in extraplanar diffuse ionized gas (eDIG) in a sample of eight nearby (17-52 Mpc) edge-on disk galaxies observed with the Multi Unit Spectroscopic Explorer (MUSE).}
         {We constructed a model for the photoionization of eDIG clouds and the propagation of ionizing radiation through the eDIG using subsequent runs of \textsc{Cloudy} photoionization code. Our model includes radiation originating both from midplane OB stars and in situ evolved stars and its dilution and processing as it propagates in the eDIG.}
         {We fit the model to the data using the vertical line ratio profiles of our sample galaxies, and find that while the ionization by in situ evolved stars is insignificant for most of the galaxies in our sample, it may be able to explain the enhanced high-ionization lines in the eDIG of the green valley galaxy ESO~544-27.}
         {Our results show that while leaking radiation from midplane H\,\textsc{ii} regions is the primary ionization source for eDIG, in situ evolved stars can play a significant part in ionizing extraplanar gas in galaxies with low star forming rates.}


\maketitle

\section{Introduction}

\begin{table*}
  \caption{Basic properties of our galaxy sample.}
  \label{tab:stats}
  \centering
  \begin{tabular}{lcccccccc}
    \hline\hline
    ID & PA & \emph{d} & $r_{25}$ & $h_{z\text{t}}$ & $h_{z\text{T}}$ & log$(M_\mathrm{t}/M_\odot)$ & log$(M_\mathrm{T}/M_{\odot})$ & $\nu_{\text{c}}$ \\

    & ($^{\circ}$) & (Mpc) & (kpc) & (kpc) & (kpc) & & & (km s$^{-1}$) \\ 
    \hline
    ESO~157-49 & 30.4 & 17.3 & 3.37 & 0.10 & 0.45 & 9.40 & 9.08 & 107 \\
    ESO~443-21 & 160.8 & 52.2 & 9.34 & 0.25 & 0.51 & 10.09 & 10.40 & 196 \\
    ESO~469-15 & 149.2 & 28.3 & 7.67 & 0.14 & 0.67 & 9.45 & 9.29 & 83 \\
    ESO~544-27 & 153.3 & 45.9 & 10.35 & 0.16 & 0.67 & 9.88 & 9.55 & 129 \\
    IC~217 & 35.7 & 21.1 & 6.13 & 0.15 & 0.54 & 9.39 & 9.24 & 115 \\
    IC~1553 & 15.0 & 36.5 & 7.17 & 0.16 & 0.48 & 9.94 & 9.60 & 142 \\
    PGC~28308 & 125.2 & 43.1 & 12.52 & 0.21 & 1.09 & 10.26 & 9.56 & 130 \\
    PGC~30591 & 169.2 & 35.5 & 10.01 & 0.15 & 0.62 & 9.60 & 9.47 & 97 \\
    \hline
  \end{tabular}
  \tablefoot{Position angle (PA) values from \cite{salo2015s4g}. Distance estimates (\emph{d}; \citealt{tully2008dist,tully2016dist}) are the same as used in \cite{comeron2018thick, comeron2019thick}. Isophotal 25 mag arcsec$^{-2}$ radius in the $B$-band ($r_{25}$) from HyperLEDA. Thin disk scale height ($h_{z\text{t}}$) and mass ($M_{\mathrm{t}}$) and thick disk scale height ($h_{z\text{T}}$) and mass ($M_\mathrm{T}$) from \cite{comeron2018thick}. Circular velocities ($\nu_{\text{c}}$) from \cite{comeron2019thick}.}
\end{table*}

Diffuse ionized gas (DIG) is ubiquitous in late-type galaxies \cite[e.g.,][]{levy2019edge}. Along with H\,\textsc{ii} regions it constitutes the H$\alpha$-bright star-forming disk visible in narrow-band imaging and integral field unit (IFU) spectroscopy. DIG also extends beyond the classical star-forming disk, to great radial distances and far above the midplane \cite[e.g.,][]{dettmar1990dig, williams1993virgo, rand1996diginedge, watkins2018m51,rautio2022}.

The ionization of diffuse gas is mainly attributed to leaking radiation of massive young OB stars from midplane H\,\textsc{ii} regions \citep{haffner2009leaky,weber2019leaky}. Leaky H\,\textsc{ii} regions have been shown to be the primary source of ionization for extraplanar DIG (eDIG) at great heights above the midplane \citep{levy2019edge, rautio2022}. However, the emission spectrum of eDIG exhibits unique features such as the enhancement of [O\,\textsc{iii}]$_{\lambda5007}$ emission line (hereafter [O\,\textsc{iii}]) that are difficult to explain without additional sources of ionization \citep[e.g.,][]{rand1998digspec2, collins2001dig, wood2004hard}.

Many alternative sources of ionization for the eDIG have been investigated in addition to leaky H\,\textsc{ii} regions. Among these alternative sources the most promise is shown by shocks \citep{dopita1995shocks, collins2001dig,ho2014shocks, dirks2023sb} and hot low-mass evolved stars \citep[HOLMES][]{flores-fajardo2011holmes, alberts2011evolution, zhang2017manga, lacerda2018ew, rautio2022}. HOLMES are also called post asymptotic giant branch stars (post-AGB), as they include all stars in subsequent stages of evolution of the asymptotic giant branch, such as white dwarfs. They are much less luminous than the bright OB stars, but also much hotter, resulting in a harder ionizing spectrum, potentially explaining the enhancement of high ionization lines such as [O\,\textsc{iii}]. While the total ionizing photon flux of HOLMES is several orders of magnitude less than that of OB stars, HOLMES are present in situ in the eDIG allowing a much greater portion of their radiation reach the eDIG \citep{stasinska2008old}.

We investigated the sources of eDIG ionization in a sample of five low-mass galaxies (ESO~157-49, ESO~469-15, ESO~544-27, IC~217, and IC~1553) in \citet[hereafter \mbox{Paper~I}]{rautio2022}, using Multi Unit Spectroscopic Explorer (MUSE) integral field spectroscopy and narrow-band imaging. While we found the primary ionization source for eDIG to be OB star radiation across the sample, we also found regions consistent with OB--HOLMES and OB--shock ionization in all of the sample galaxies. Our analysis consisted of photometry and spatially resolved emission-line diagnostics, and was mostly qualitative in nature. Here we extend the analysis to three more edge-on galaxies observed with MUSE (ESO~443-21, PGC~28308, and PGC~30591), and quantify the significance of HOLMES to the eDIG ionization by modeling its photoionization with version 23.00 of the \textsc{Cloudy} code \citep{cloudy23}.

Modeling the ionization of diffuse extended gas such as the eDIG poses unique challenges due to the large range of spatial scales involved. The eDIG can extend kiloparsecs both in the disk-plane and perpendicular to it, and at the same time exhibit porous and turbulent structure at the parsec scale as revealed by both observations \citep{dettmar1990dig, rand1990wim, collins2000dig, miller2003twodisk, lopez-coba2019outflows}, and hydrodynamical simulations {\citep{wood2010sim, barnes2014sim, barnes2015hds, vandenbroucke2018sim, kado-fon2020sim}}. The fine structure of the eDIG may include chimneys, holes and clouds, and it is mixed with neutral gas, as indicated by the presence of neutral emission lines such as [O\,\textsc{i}]$_{\lambda6300}$ (hereafter [O\,\textsc{i}]) in eDIG spectra \citepalias[e.g.,][]{rautio2022}. However, focusing on integrated properties allows us to create a model for the photoionization of eDIG that is able to reproduce the observed line ratios in our sample.

We describe the sample, the data, and the data analysis in Sect. \ref{sec:obs}. In Sect. \ref{sec:models}, we present our model for the eDIG photoionization. We describe the best fit models for our sample galaxies and discuss the implications in Sect. \ref{sec:results}.  In Sect. \ref{sec:disc}, we discuss the caveats in our modeling approach and the nature of the eDIG. We finish with a summary and conclusions in Sect. \ref{sec:sum}.

\section{Observations and data analysis}

\label{sec:obs}
\subsection{The sample}
Our sample consists of eight edge-on galaxies observed by \cite{comeron2019thick} with MUSE at the Very Large Telescope (VLT) of the European Southern Observatory (ESO), at Paranal observatory, Chile. The galaxies come from a subsample of the 70 edge-on disk sample of \cite{comeron2012breaks}, that itself was derived from the Spitzer Survey of Stellar Structure in Galaxies (S$^4$G; \citealt{sheth2010s4g}), selected to be observable from the Southern Hemisphere and to have an $r_{25} < 60\arcsec$ (as given in the HyperLEDA\footnote{\url{http://leda.univ-lyon1.fr/}} database; \citealt{makarov2014hyperleda}). This selection was done so that the region between the center of the target and its edge could be covered in a single MUSE pointing. The galaxies have relatively low masses, with stellar masses ranging from $4\times10^9 M_\odot$ to $4\times10^{10} M_\odot$. Most of them also have clear separate thin and thick disks, while for two (ESO~443-21 and PGC~30591) the disk structure is less clear \citep{comeron2018thick}. Some properties, including the masses and scale heights of thin and thick disks, of the sample galaxies can be found in Table \ref{tab:stats}.

\subsection{MUSE observations}
For the MUSE observations, four 2624 s on-target exposures were taken per galaxy (three for IC~217), centered either on a single point on one side of the galaxy, or in the case of galaxies with smaller angular size (ESO~157-49 and IC~1553), on the center of the galaxy, with 90\degree~rotations between exposures. The MUSE pipeline \citep{weilbacher2012pipeline} was used within the Reflex environment \citep{freudling2013reflex} to reduce the data, while the Zurich Atmosphere Purge (ZAP; \citealt{soto2016zap}) was used to clean the combined datacubes of sky residuals. To produce bins with S/N $\sim$ 50 in H$\alpha$ the datacubes were tessellated with the Voronoi binning code by \cite{cappellari2003voronoi}. Intensities of H$\alpha$, H$\beta$, [O\,\textsc{iii}], [O\,\textsc{i}], [N\,\textsc{ii}]$_{\lambda6583}$, and [S\,\textsc{ii}]$_{\lambda\lambda6717/31}$ (hereafter [N\,\textsc{ii}], and [S\,\textsc{ii}]) emission-lines were obtained with the Python version of the Penalized Pixel-Fitting code (pPXF; \citealt{cappellari2004ppxf}). The MUSE observations, data reduction, Voronoi binning, and the pPXF processing are described in detail in \cite{comeron2019thick}.

\subsection{H$\alpha$ photometry}

For the star forming rates (SFR) and eDIG vertical profile parameters of most of our sample galaxies, we used the \citetalias{rautio2022} H$\alpha$ photometry obtained from narrow-band data. However, we did not obtain H$\alpha$ narrow-band imaging for the three galaxies that were not investigated in \citetalias{rautio2022} (ESO~443-21, PGC~28308, and PGC~30591). Instead we used MUSE data to obtain H$\alpha$ photometry for these three galaxies by mapping the Voronoi-binned H$\alpha$ emission-line intensity back to the MUSE grid.

To obtain the midplane H\,\textsc{ii}-region scale height ($h_{z\text{H}\,\textsc{ii}}$) and the eDIG scale height ($h_{z\text{eDIG}}$) we used the same process as described in detail in \citetalias{rautio2022}. Following is a brief explanation of the process: We first fit a line-of-sight integrated two-exponential disk model to the vertical profiles measured from the Voronoi-binned H$\alpha$ maps, and obtained $h_{z\text{H}\,\textsc{ii}}$ from the scale height of the thinner component. Then we fit the line-of-sight integrated two-exponential disk model to the H$\alpha$ maps again, this time masking out the midplane up to $z = h_{z\text{H}\,\textsc{ii}}$, and obtained $h_{z\text{eDIG}}$ from the scale height of the thicker component.

The main limitation of the MUSE data compared to the narrow-band data is the comparatively small field of view (FoV) of MUSE: most of the sample galaxies have a larger angular size parallel to the disk plane than the FoV of MUSE. This was a more significant issue for SFR measurement than the measurement of vertical profile averaged over the galaxy radius, as we obtained the SFRs from the integrated H$\alpha$ intensities. To obtain SFRs from MUSE data that are comparable to the ones we obtained from narrow-band data in \citetalias{rautio2022}, we mirrored the H$\alpha$ maps over the centers of the galaxies and replaced the missing data in each galaxy with its mirror over the $z$-axis. We then measured the integrated H$\alpha$ flux of these mirrored H$\alpha$ maps, correcting for extinction using color excess derived from Balmer decrement as

\begin{equation}
  E(B-V) = \frac{2.5}{k(\mathrm{H}\beta) - k(\mathrm{H}\alpha)} \log{\left(\frac{\mathrm{H}\alpha/\mathrm{H}\beta}{2.86}\right)},
\end{equation}

\noindent where $E(B-V)$ is the color excess, and $k$ are obtained from the \cite{calzetti2000extinction} extinction law as

\begin{equation}
  \begin{array}{l}
    k(\mathrm{H}\alpha) = 2.659 (-1.857 + 1.040/\lambda_{\mathrm{H}\alpha}) + R_V, \quad \mathrm{and} \\
    \begin{aligned}
    k(\mathrm{H}\beta)\, =\: & 2.659 (-2.156 + 1.509/\lambda_{\mathrm{H}\beta} \\
    &- 0.198/\lambda_{\mathrm{H}\beta}^2 + 0.011/\lambda_{\mathrm{H}\beta}^3) + R_V,
    \end{aligned}
  \end{array}
\end{equation}

\noindent where $R_V = 4.5$ \citep{fischera2005rv}, and $\lambda_{\mathrm{H}\alpha}$ and $\lambda_{\mathrm{H}\beta}$ are the wavelengths of H$\alpha$ and H$\beta$. From the extinction-corrected H$\alpha$ flux we calculated the H$\alpha$ luminosity ($L_{\mathrm{H}\alpha}$) using distances from \cite{tully2008dist,tully2016dist}, and obtained the SFRs using the \cite{hao2011sfr} and the \cite{murphy2011sfr} calibrations as

\begin{equation}
  \log{\left(\frac{\mathrm{SFR}}{M_\odot \mathrm{yr}^{-1}}\right)} = \log{\left(\frac{L_{\mathrm{H}\alpha}}{\mathrm{erg}s^{-1}}\right)} - 41.27.
\end{equation}

The SFRs and eDIG vertical profiles for ESO~443-21, PGC~28308, and PGC~30591, were obtained with incomplete spatial coverage due to the limited MUSE FoV. As such, the uncertainty of these measurements is high. To evaluate this uncertainty, we measured SFRs from the MUSE data in the same manner for the five galaxies for which we did have narrow-band data, and compared the extrapolated MUSE SFRs to the narrow-band SFRs. The largest difference between the MUSE SFRs and the narrow-band SFRs we found was MUSE SFR $=0.18 M_\odot$yr$^{-1}$ for IC~217 compared to narrow-band SFR $=0.25 M_\odot$yr$^{-1}$ \citep{rautio2022}. The star-forming properties of our sample galaxies are gathered in Table \ref{tab:sfr}.

\begin{table}
  \caption{Star-forming properties of our galaxy sample.}
  \label{tab:sfr}
  \centering
  \begin{tabular}{lcccc}
    \hline\hline
    ID & SFR & $h_{z\text{H}\,\textsc{ii}}$ & $h_{z\text{eDIG}}$ & Source\\

    & ($M_\odot$yr$^{-1}$) & (kpc) & (kpc) & \\ 
    \hline
    ESO~157-49 & 0.19 & 0.20 & 0.60 & \citetalias{rautio2022} \\
    ESO~443-21 & 2.40 & 0.66 & 1.08 & This paper \\
    ESO~469-15 & 0.20 & 0.13 & 1.32 & \citetalias{rautio2022} \\
    ESO~544-27 & 0.13 & 0.22 & 0.61 & \citetalias{rautio2022} \\
    IC~217 & 0.25 & 0.22 & 1.12 & \citetalias{rautio2022} \\
    IC~1553 & 0.96 & 0.25 & 1.39 &  \citetalias{rautio2022} \\
    PGC~28308 & 0.51 & 0.36 & 1.08 & This paper \\
    PGC~30591 & 0.29 & 0.24 & 0.91 & This paper\\
    \hline
  \end{tabular}
  \tablefoot{$h_{z\text{H}\,\textsc{ii}}$  and $h_{z\text{eDIG}}$ are the scale heights of H\,\textsc{ii} regions and the eDIG, respectively.}
\end{table}

\subsection{Metallicity}

We produced Voronoi-binned metallicity maps for all galaxies in our sample using the emission-line intensities obtained with pPXF. We used the calibration of \cite{dopita2016metals} to do this. We define metallicity here with the logarithmic oxygen--hydrogen ratio $Z = 12+\log(\mathrm{O}/\mathrm{H})$\footnote{The Sun has $12+\log(\mathrm{O}/\mathrm{H}) \approx 8.7$}.

We measured vertical metallicity profiles from these Voronoi-binned metallicity maps by averaging the metallicity at each height. We only used bins with OB star dominated ionization in the averages, as \cite{dopita2016metals} calibration was derived for H\,\textsc{ii}-regions where ionization is caused by star formation. We determined the ionization source of the bins using the $\eta$-parameter \citep{erroz-ferrer2019eta}. The $\eta$-parameter is defined as a distance to the \cite{kewley2001starburst} and \cite{kauffmann2003sfline} starburst lines in the [N{\sc ii}]/H$\alpha$ against [O{\sc iii}]/H$\beta$ Baldwin-Phillips-Terlevich (BPT; \citealt{baldwin1981spectra}) diagram. Ionized gas having $\eta < -0.5$ indicates that it is located below the \cite{kauffmann2003sfline} line on the BPT diagram, and its ionization is dominated by radiation from OB stars, $\eta > 0.5$ indicates that it is located above the \cite{kewley2001starburst} line on the BPT diagram, and its ionization is dominated by other ionization sources, while $ -0.5 < \eta < 0.5$ indicates mixed ionization between the two starburst lines.

In all of the galaxies, the measured metallicity profile exhibits a trend where it first decays exponentially as $z$ increases, but after some height the metallicity either remains roughly constant or begins increasing again. The apparent enhancement of metallicity in the eDIG at large $z$ is likely not a physical effect, but rather results from the gas density of the eDIG dropping below the limits where \cite{dopita2016metals} calibration holds. We adopt a constant metallicity for the eDIG, and fit to the data a profile of the form

\begin{equation}
  Z(z) = (Z_0 - Z_{\mathrm{eDIG}}) \mathrm{e}^{-z/h_{z\mathrm{O}/\mathrm{H}}} + Z_{\mathrm{eDIG}},
  \label{eq:metals}
\end{equation}
  
\noindent where $Z_0$ is the metallicity at the midplane, $Z_{\mathrm{eDIG}}$ is the metallicity of the eDIG, and $h_{z\mathrm{O}/\mathrm{H}}$ is the scale height for the exponential decay of the metallicity between the midplane and the eDIG. The metallicity parameters for our sample are shown in Table \ref{tab:metal}

\begin{table}
  \caption{Metallicity parameters for our galaxy sample.}
  \label{tab:metal}
  \centering
  \begin{tabular}{lccc}
    \hline\hline
    ID & $Z_0$ & $Z_\mathrm{eDIG}$ & $h_{z\mathrm{O}/\mathrm{H}}$ \\

    & & & (kpc) \\ 
    \hline
    ESO~157-49 & 8.51 & 8.35 & 0.31 \\
    ESO~443-21 & 8.57 & 8.38 & 0.49 \\
    ESO~469-15 & 8.47 & 8.22 & 0.73 \\
    ESO~544-27 & 8.49 & 8.33 & 2.19 \\
    IC~217 & 8.14 & 7.99 & 0.25 \\
    IC~1553 & 8.44 & 8.24 & 0.84 \\
    PGC~28308 & 8.64 & 8.37 & 2.16 \\
    PGC~30591 & 8.39 & 8.26 & 0.24 \\
    \hline
  \end{tabular}
  \tablefoot{$Z = 12+\log(\mathrm{O}/\mathrm{H})$. See Eq. \ref{eq:metals}}
\end{table}

\section{Model for the photoionization of extraplanar diffuse ionized gas}
\label{sec:models}

\subsection{Modeling scenario}
\begin{figure}
  \centering
  \includegraphics[width=0.5\textwidth]{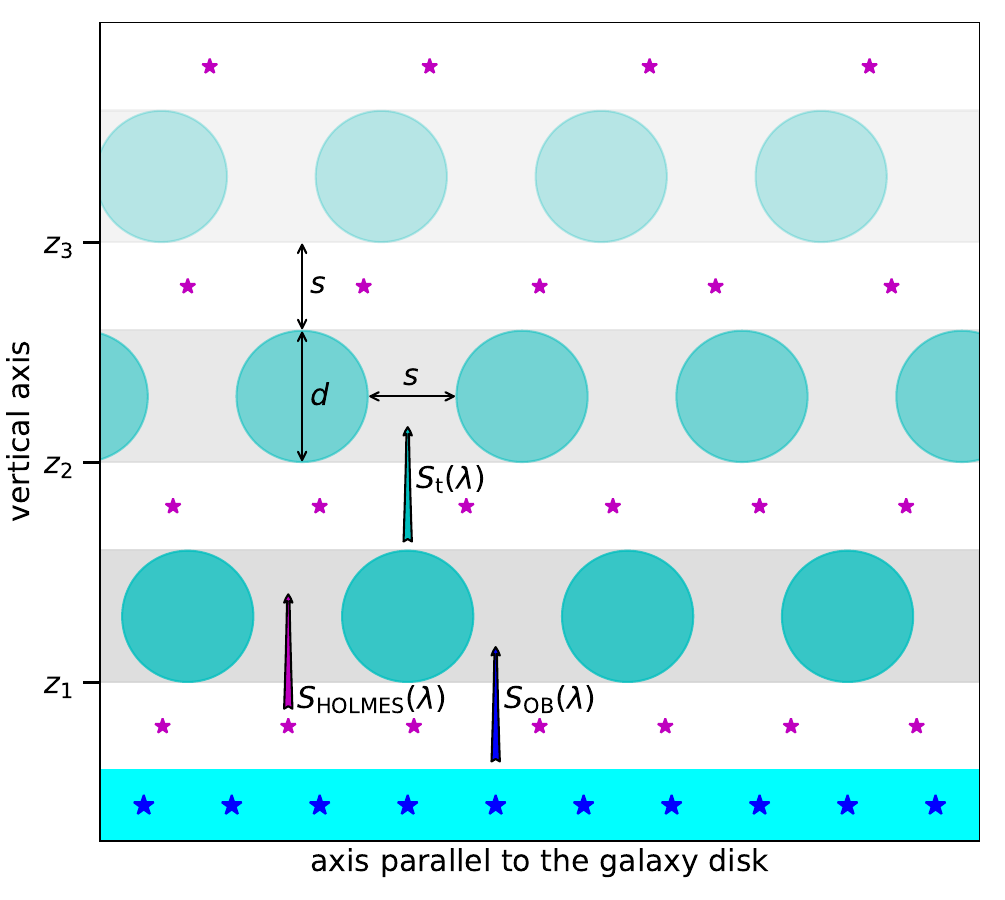}
  \caption{Schematic of our model. The cyan shaded area indicates the star-forming disk and the large blue stars represent the OB stars within it. The gray shaded areas indicate the eDIG cloud layers, and the filled cyan circles show the eDIG clouds within them. The small purple stars represent the HOLMES between the cloud layers. The colored arrows indicate the source radiation fields of leaking midplane H\,\textsc{ii} regions ($S_\mathrm{OB}(\lambda)$; blue), HOLMES ($S_\mathrm{HOLMES}(\lambda)$; magenta), and transmitted radiation ($S_\mathrm{t}(\lambda)$; cyan). Black arrows indicate the cloud diameter ($d$) and the layer separation ($s$).}
  \label{fig:sche}
\end{figure}

We investigated here a simplified scenario where spherical clouds of diffuse gas, arranged in equidistant layers with increasing height over the midplane ($z$) and decreasing gas density, are ionized by OB star radiation originating from the midplane and HOLMES radiation originating from between the layers. Figure \ref{fig:sche} shows a schematic view of our model. The radiation field of leaking midplane H\,\textsc{ii} regions $S_\mathrm{OB}(\lambda)$ and the radiation field of in situ HOLMES $S_\mathrm{HOLMES}(\lambda)$, as well as the radiation field transmitted through the clouds $S_\mathrm{t}(\lambda)$, are shown in the schematic with colored arrows. We use ``transmitted radiation'' here to describe the sum of attenuated incident radiation and the emission of the cloud itself, following \textsc{Cloudy} documentation nomenclature. The transmitted radiation field contains both the processed radiation field of midplane OB stars and the processed radiation field of HOLMES. 

The individual eDIG clouds in our model have a constant gas density determined by their distance from the midplane, following an exponential decay.  Each cloud layer also has a constant metallicity determined by Eq. \ref{eq:metals}. Our code also models the propagation of radiation between the layers taking into account the filling factor of the eDIG and average geometric dilution with the assumption that the clouds are located above the center of the galaxy disk. The determination of these parameters is explained Sect. \ref{sec:geo}, and Sect. \ref{sec:fit}

We used version 23.00 of \textsc{Cloudy} photoionization code \citep{cloudy23} to model the ionization of each cloud layer. \textsc{Cloudy} is a 1-dimensional (1D) spectral synthesis code which inputs a radiation field into a cloud of interstellar gas and calculates the resulting physical condition in the cloud as well as the emitted and transmitted spectra in a plane-parallel or spherically symmetric geometry. In our model \textsc{Cloudy} is ran once for each cloud layer using plane-parallel geometry and the radiation field leaking from midplane H{\sc ii} regions incident on the cloud surface $I_\mathrm{OB}(\lambda)$, the incident radiation field from inter-layer HOLMES $I_\mathrm{HOLMES}(\lambda)$, and the incident radiation field transmitted through clouds in the lower layers $I_\mathrm{t}(\lambda)$ is input, simulating an ``average'' eDIG cloud for that layer. We use $I(\lambda)$ to denote the diluted radiation fields incident on the clouds and $S(\lambda)$ to denote the undiluted source radiation fields\footnote{We use $I(\lambda)$ and $S(\lambda)$ here for brevity. In practice we input to \textsc{Cloudy} the spectrum and the flux of hydrogen ionizing photons hitting the surface of the cloud $\Phi = \int_{0 \textrm{Å}}^{911.3 \textrm{Å}} I(\lambda) \,d\lambda$.}.

{Our model is 1-dimensional, predicting the evolution of line ratios in the eDIG only as a function of $z$. The eDIG of real galaxies varies also along the disk plane. To take this into account we considered adopting a more complex modeling scenario where the model varies in coordinates $(r, z)$, where $r$ is the radial direction along the galaxy disk. However, we opted not to adopt this scenario due to several issues with it. Firstly, as we show in \citetalias{rautio2022}, the line-of-sight integrated radial H$\alpha$ profiles of our sample galaxies are complex and cannot be decomposed into simple exponential disks. This would mean the radial dependence of OB star flux and gas properties would need to be left as free parameters. Secondly, as \textsc{Cloudy} is used in our model to compute $S_\mathrm{t}(\lambda)$ and \textsc{Cloudy} is a 1D code, the incident direction of radiation transmitted through the eDIG clouds is lost and thus the treatment of transmitted radiation would not be realistic in a 2-dimensional model. Thirdly, in a 2-dimensional model we would need to model individual clouds in each cloud layer separately, drastically increasing the computational load. Because of these reasons, we chose our 1D modeling approach. We consider the limitations of our approach throughout the following Sections.}    

\subsection{Geometry}
\label{sec:geo}
Before going into details about the ionization sources of our model and the parameters that we used for the \textsc{Cloudy} runs, let us first examine the geometry of our scenario. The eDIG clouds in our model are arranged in grids within their layers so that the distance between a cloud and its four nearest neighbors within the layer is equal to the distance between cloud layers. In our scenario this distance $s$ is related to the volume filling factor ($f_V$) as follows:

\begin{equation}
  f_V = \frac{\pi}{6} \frac{d_\mathrm{cloud}^3}{(d_\mathrm{cloud} + s)^3},
\end{equation}

\noindent where $d_\mathrm{cloud} = 100$ pc is the diameter of the clouds, which is equal to the layer thickness. From this follows that the cloud cross-sections cover a fraction

\begin{equation}
f_S = \frac{\pi}{4} \frac{d_\mathrm{cloud}^2}{(d_\mathrm{cloud} + s)^2} = \sqrt[3]{\frac{9\pi}{16}} f_V^{\frac{2}{3}},
\end{equation}

\noindent of  each layer, where $f_S$ is the surface filling factor. The cloud layers are arranged so that the lower boundary of the first layer is at height $z_1$, and the following layers are at heights $z_j = (j-1)(s+d_{\mathrm{cloud}}) + z_1$, where $j$ is the layer number.

As radiation propagates between the cloud layers, a fraction of the radiation field intensity is lost to ionizing the eDIG within the clouds and to geometric dilution. This affects the intensity of a radiation field incident to a cloud layer in the following way:

\begin{equation}
  I_i(\lambda) = g_{ij} \times (1-f_S)^{|j-i|} \times S_j(\lambda),
  \label{eq:phi}
\end{equation}

\noindent where $S_j(\lambda)$ is the source radiation field at layer $j$, $I_i(\lambda)$ is the contribution to the incident radiation field at layer $i$, $(1-f_S)^{|j-i|}$ is the fraction of photons from layer $j$ that reach layer $i$ without striking any clouds, and $g_{ij}$ is the geometric dilution factor between layers $i$ and $j$. This applies for all $i$ and $j$, regardless of the direction or source of the radiation.

The radii of the H$\alpha$ disks of our galaxies are of similar order as the heights over midplane that eDIG reaches for our sample. This means that geometric dilution of the radiation is significant for our model, but not as strong as it is for a spherically symmetric system. We approximate the geometric dilution in our model as the dilution of the radiation field along the $z$-axis of a uniform disk source with radius $R$, which corresponds to the solid angle subtended by a disk with radius $R$ at a perpendicular distance $dz$ above the disk center

\begin{equation}
  \Omega = 2\pi \left( 1 - \frac{1}{\sqrt{1 + R^2 / dz^2}} \right).
  \label{eq:sa}
\end{equation}

\noindent Normalizing Eq. \ref{eq:sa} so that it goes to unity in a plane-parallel scenario ($R \rightarrow \infty$), we obtain

\begin{equation}
  g_{ij} = 1 - \frac{1}{\sqrt{1 + R^2 / (z_j - z_i)^2}},
\end{equation}

\noindent as the geometric dilution factor.

\subsection{Leaking radiation from H{~\sc ii} regions}
\label{sec:ob}
To model the radiation of the midplane population of OB stars, we used \textsc{Starburst99} \citep{leitherer1999sb99} stellar population synthesis code. We adopted the \cite{kroupa2001imf} initial mass function (IMF), and the Geneva stellar tracks with solar metallicity and no rotation \citep{ekstrom2012tracks}. For the stellar model atmospheres, we used those of \cite{pauldrach2001atmo} and \cite{hillier1998atmo}, implemented into \textsc{Starburst99} by \cite{smith2002atmo}. We assumed constant star formation and extracted the radiation field at $10^7$ yr age, when it is stabilized.

Most midplane OB stars are located within H\,{\sc ii} regions, and it is likely that a significant portion of the flux that escapes midplane is processed by propagation through density bounded H\,{\sc ii} regions, hardening the radiation field above the midplane. To simulate this hardening of the radiation field, we performed preliminary \textsc{Cloudy} runs of a density bounded H\,{\sc ii} regions for each galaxy in our sample, inputting the spectrum generated by \textsc{Starburst99}, and taking as output the transmitted spectra from \textsc{Cloudy}. We used as the H\,{\sc ii} regions 0.5 pc thick clouds with densities of 100 cm$^{-3}$, illuminated by a radiation fields with hydrogen-ionizing photon surface fluxes $\Phi = 10^{13}$ cm$^{-2}$ s$^{-1}$. These \textsc{Cloudy} parameters ensured a fully ionized, density bounded clouds with escape fractions that are in the high end of typical values derived for real H\,{\sc ii} regions ($e_{\mathrm{H}\,\textsc{ ii}} = 0.6$; \citealt{db2021ef, teh2023ef}). We chose to simulate a very ``leaky'' H\,{\sc ii} regions as they would be overpresented among the sources of the ionizing photons that reach the eDIG. Additionally, majority of radiation leakage in real H\,\textsc{ii} regions most likely happens through empty holes, in which case the radiation field of the OB stars is not processed at all \citep{kimm2022hii}. Thus, our model likely overestimates the hardening of OB star radiation field by partial absorption in the surrounding H\,\textsc{ii} regions. This means that if the leaking OB star radiation in our model fails to reproduce the observed high-ionization line intensities in the eDIG, additional ionization sources must be present. We set the metallicities of the H\,{\sc ii} regions according to $Z_0$ of the modeled galaxy.

We scaled the transmitted fields from the preliminary \textsc{Cloudy} runs depending on the galaxy we were modeling to obtain the source radiation fields of the leaking midplane radiation $S_{\mathrm{OB}}(\lambda)$. By following Eq. \ref{eq:phi} we then obtained the incident radiation fields of the leaking midplane radiation for each cloud layer as

\begin{equation}
  I_{\mathrm{OB},i}(\lambda) = g_{i1} \times (1-f_S)^{|1-i|} \times S_{\mathrm{OB}}(\lambda), \quad i = 1,\dotsc,N,
\end{equation}

\noindent where $N$ is the number of layers.

\subsection{In situ radiation from evolved stars}
\label{sec:holmes}

\textsc{Starburst99} does not include the radiation field of post-AGB stars, and as such is not well suited to modeling the spectra of HOLMES. We instead used \textsc{P\'egase}.3 spectral evolution synthesis code \citep{fioc2019pegase} to produce the spectra of HOLMES for our model. We follow the approach of \cite{flores-fajardo2011holmes}, and take the spectrum of a coeval population of stars at an age of 10 Gyr as the spectral energy distribution of HOLMES. Instantaneous burst produces a realistic spectrum, as the integrated radiation field from all HOLMES in an old stellar population (older than 100 Myr) is nearly independent of its star formation history. We used \cite{kroupa1993imf} IMF and solar metallicity for the \textsc{P\'egase}.3 run.

HOLMES are distributed throughout the thick disk and halo, unlike the OB stars that are primarily confined to the midplane, so HOLMES are inserted between the cloud layers in our model. Our model includes two HOLMES components, one follows the vertical profile of the thick disk, while the other is a flat background component truncated at the height of the final cloud layer with 10\% of the mass of the thick disk component, representing the halo. Both components use the same \textsc{P\'egase}.3 spectrum. We scaled the intensity of the spectrum according to the thick disk mass of the galaxy we were modeling, and distributed it over the HOLMES components according to their vertical profiles. This gave us the radiation fields originating from each inter-layer HOLMES layer $S_{\mathrm{HOLMES},j}(\lambda)$. Then for each cloud layer $i$ we calculated

\begin{equation}
  I_{\mathrm{HOLMES},i}(\lambda) = \sum^{N}_{j=1} g_{ij} (1-f_S)^{|j-i|} S_{\mathrm{HOLMES},j}(\lambda), \; i = 1,\dotsc,N,
\end{equation}

\noindent where $I_{\mathrm{HOLMES},i}(\lambda)$ is the sum of all the inter-layer HOLMES layer radiation fields incident at layer $i$.

\subsection{Radiation transmitted through the eDIG clouds}
\label{sec:trans}
The final component of the incident radiation field for each cloud layer in our model is the radiation transmitted through clouds in the lower layers. To obtain the transmitted radiation field incident at a given layer $I_{\mathrm{t},i}(\lambda)$, we took the transmitted spectra given by \textsc{Cloudy} for each lower layer $S_{\mathrm{t},j}(\lambda)$ and computed the propagation of radiation as follows:

\begin{equation}
  I_{\mathrm{t},i}(\lambda) = \sum_{j < i} g_{ij} (1-f_S)^{|j-i|} f_S S_{\mathrm{t},j}(\lambda), \quad i = 1,\dotsc,N,
  \label{eq:trans}
\end{equation}

\noindent where

\begin{equation}
  S_{\mathrm{t},j}(\lambda) = C_j \circ [I_{\mathrm{HOLMES},j}(\lambda) + I_{\mathrm{OB},j}(\lambda) + I_{\mathrm{t},j}(\lambda)], \;\; j = 1,\dotsc,N,
\end{equation}

\noindent and $C_j \circ$ is an operator representing the processing and hardening of the radiation field caused by a transmission through a cloud of layer $j$.

Our treatment of the transmitted radiation did not take into account the directionality of the HOLMES radiation. That is, the fact that HOLMES radiation originating from higher layers should be transmitted mostly towards the lower layers rather than back upwards. As \textsc{Cloudy} does not differentiate between the origin of the photons in the transmitted spectrum, a more realistic approach taking into account the differing directions of the radiation field components is difficult to implement. However, this does not affect our results, as while the harder radiation field from HOLMES can have a significant effect on the ionization state of the eDIG, the HOLMES radiation makes up such a small fraction of the total radiation field that the exact distribution of the HOLMES radiation is not important. This is not an issue for the OB star radiation escaping from the midplane, as it all originates from below all of the eDIG cloud layers. We performed a more rigorous treatment that took into account the direction of the transmitted HOLMES radiation for models of two galaxies as described in Sect. \ref{sec:hcon}.

\subsection{Fitting the model to data}
\label{sec:fit}

The input parameters for the model described above are SFR, escape fraction from the midplane ($e_{\textrm{mp}}$), radius of the galaxy ($R$), mass of the thick disk ($M_\mathrm{T}$), scale height of the thick disk ($h_{z\mathrm{T}}$), the ionization parameter at the surface of the first eDIG cloud ($U_1$), the scale height of the gas density ($h_{zn}$), the metallicity parameters ($Z_0$, $Z_{\mathrm{eDIG}}$, and $h_{z\mathrm{O}/\mathrm{H}}$), the filling factor of the eDIG ($f_V$), the number of cloud layers ($N$), and the height of the first cloud layer ($z_1$). Using these parameters our code computes the input parameters for the \textsc{Cloudy} runs of each cloud layer and runs \textsc{Cloudy} to obtain the predicted eDIG emission-line intensities at the height of each cloud layer.

\begin{figure}
  \centering
  \includegraphics[width=0.5\textwidth]{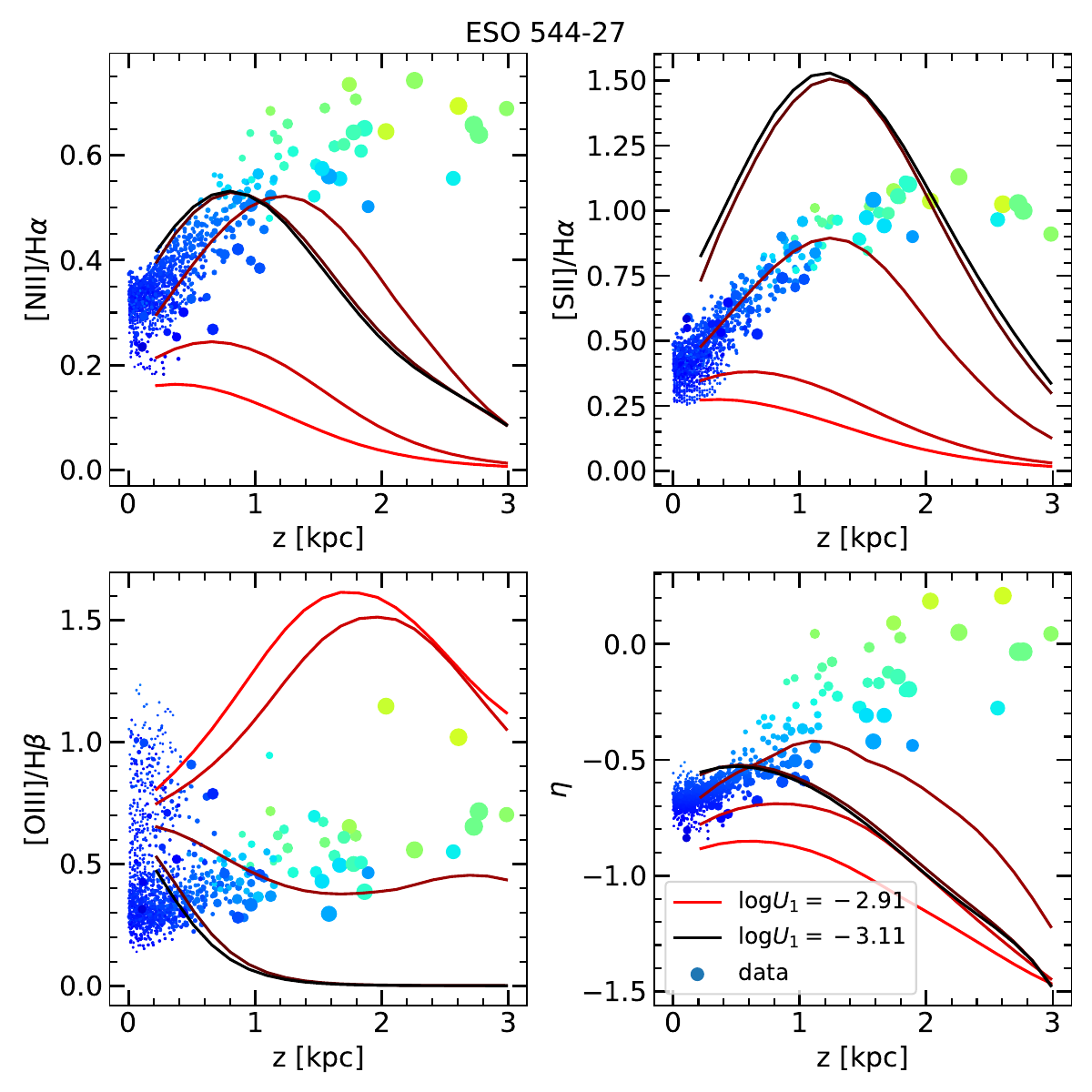}
  \caption{Line ratio and $\eta$-parameter profiles for ESO~544-27 models with varying $U_1$. The dots show the observed values, colored according to the $\eta$-parameter going from blue at low $\eta$ to green at high $\eta$. The size of each dot is relative to the size of the corresponding Voronoi bin. The curves show the model predictions with the color of the curve corresponding with the model $\log U_1$, ranging from $\log U_1 = -3.11$. (black) to $\log U_1 = -2.91$ (red) with $\log U_1 = 0.05$ steps. All the models have $f_V = 0.17$.}
  \label{fig:varyU}
\end{figure}

\begin{figure}
  \centering
  \includegraphics[width=0.5\textwidth]{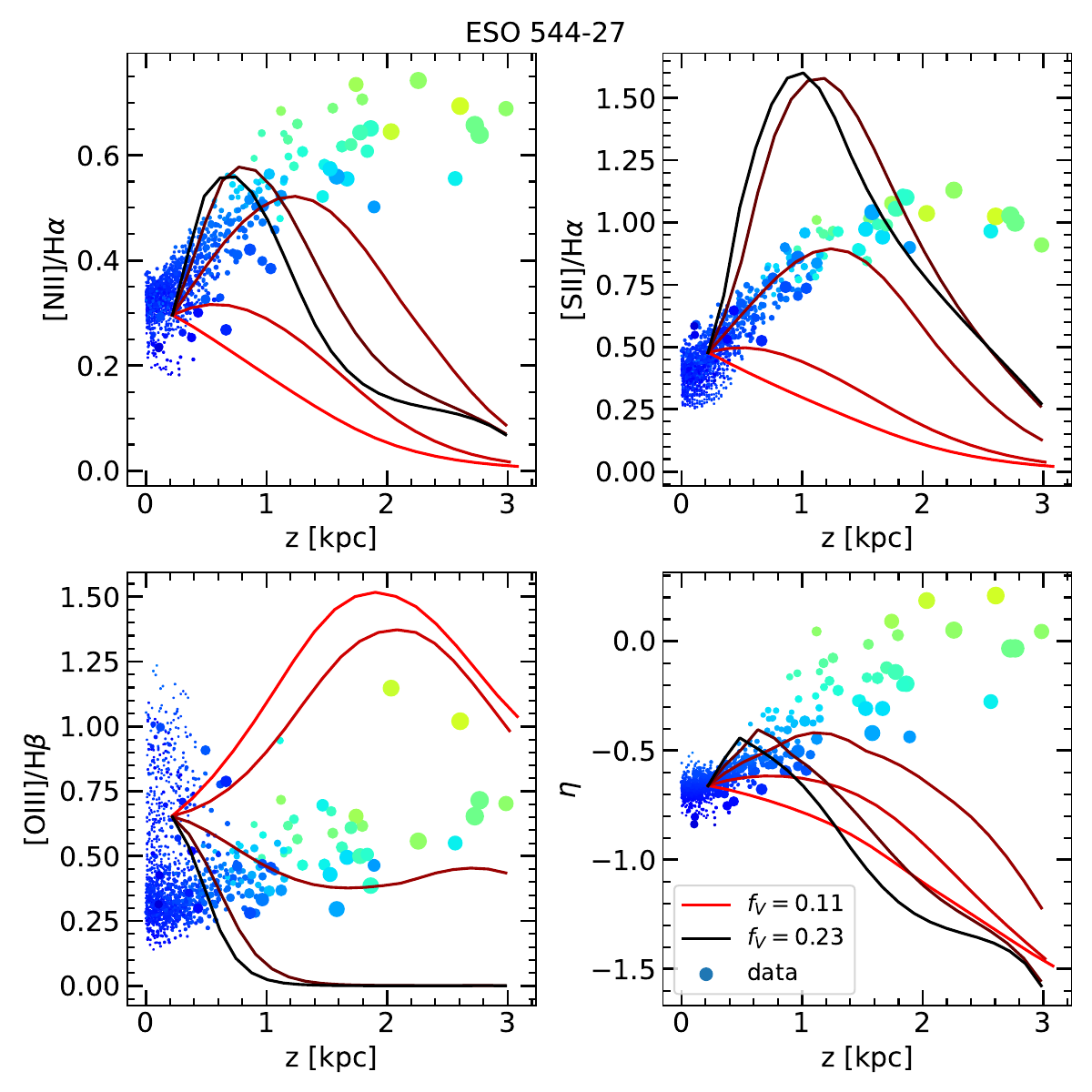}
  \caption{Line ratio and $\eta$-parameter profiles for ESO~544-27 models with varying $f_V$. The dots show the observed values, colored according to the $\eta$-parameter going from blue at low $\eta$ to green at high $\eta$. The size of each dot is relative to the size of the corresponding Voronoi bin. The curves show the model predictions with the color of the curve corresponding with the model $f_V$, ranging from $f_V = 0.11$. (red) to $f_V = 0.23$ (black) with $f_V = 0.03$ steps. All the models have $\log U_1 = -3.01$.}
  \label{fig:varyff}
\end{figure}

We obtained $M_\mathrm{T}$, $h_{z\mathrm{T}}$, SFR, $h_{z\mathrm{eDIG}}$, and the metallicity parameters from Tables \ref{tab:stats},  \ref{tab:sfr}, and \ref{tab:metal}, while for $z_1$ we used the scale height of H\,\textsc{ii} regions ($h_{z\text{H}\,\textsc{ii}}$) from Table \ref{tab:sfr}, and for $R$ we took the radius of the H$\alpha$ disk measured by eye. For the escape fraction we used $e_{\textrm{mp}} = 0.5$. This is lower than $e_{\textrm{H\,\textsc{ii}}} = 0.6$ used in our models of density bounded H\,\textsc{ii} regions as we assume that some H\,\textsc{ii} regions are fully ionization bounded, or have holes directed only along the plane of the galaxy, and thus leak no radiation to the eDIG \citep{teh2023ef}. The number of cloud layers was chosen for each model run so that the model extends to roughly equal height above the midplane as the MUSE data of the galaxy being modeled. This left $U_1$ and $f_{V}$ as free parameters. We chose these as the free parameters because it is not possible to measure them from our MUSE data, yet they are crucial in determining the observed line ratios in the eDIG. The ionization state of the first layer strongly depends on $U_1$, while the gas density, $h_{zn}$, and $f_{V}$ are the primary determinants in the absorption and processing of the radiation field that occurs as it propagates through the eDIG, which in turn determines how the ionization structure changes as a function of $z$. The effects of $U_1$ and $f_{V}$ on the model are illustrated by Figures \ref{fig:varyU} and \ref{fig:varyff}, that show line ratio and $\eta$-parameter profiles for sets of model runs for ESO~544-27 where other parameters are kept constant but $U_1$ and $f_V$, respectively, are varied. Similar figures for the effect of other parameters on our model are shown in Appendix \ref{a:vary}.

\begin{figure}
  \centering
  \includegraphics[width=0.5\textwidth]{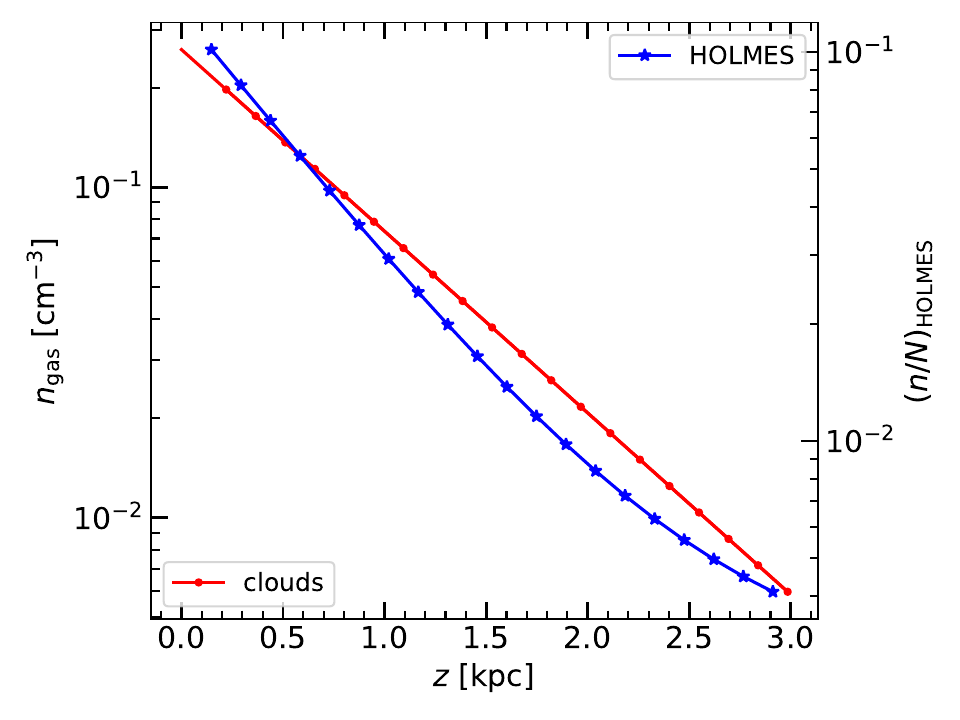}
  \caption{Density profile of gas and HOLMES for an ESO~544-27 model. Red dots indicate the cloud layers and blue stars indicate the HOLMES layers. The connecting solid curves plot the densities. For HOLMES, the relative number density defined as the fraction of HOLMES at each layer is plotted. The model parameters are $\log U_1 = -3.01$, which is used to determine the density at midplane (see text), and the scale heights of the gas density ($h_{zn} = 0.79$ kpc), and HOLMES number density ($h_{z\mathrm{T}} = 0.67$ kpc). The gas density of the clouds follows an exponential profile, while the HOLMES number density profile has an exponential component and a constant component with 10\% of the mass of the exponential component ($(n/N)_\mathrm{HOLMES} = 2.5 \times 10^{3}$).}
  \label{fig:nH}
\end{figure}

Our model uses the dimensionless ionization parameter, defined as $U_1 = \Phi_1 / (n_1 c)$, where $\Phi_1$ is the total hydrogen ionizing photon flux at the surface of the first cloud and $c$ is the speed of light, to calculate $n_1$, the gas density of the first cloud. We chose to use $U_1$ as an input parameter rather than $n_1$ because the ionization structure of a low-density gas such as eDIG is primarily determined by $U$ rather than $n$ and $\Phi$ individually. As scale height of H$\alpha$ emission is closely related to gas density scale height, we used the scale height of the disturbed eDIG ($h_{z\mathrm{eDIG}}$) to determine $h_{zn}$. Due to geometric dilution of the ionizing radiation field, the H$\alpha$ intensity declines faster than the gas density. \cite{seon2009dig} constructed two exponential gas disk models (one with $h_{zn} = 1$ kpc and one with $h_{zn} = 1.5$ kpc) for the face-on spiral M~51, and let the disks be ionized by radiation leaking from observed midplane H{\sc ii} regions. Their modeling produced vertical H$\alpha$ profiles consisting of two exponential components with the thicker component having a scale height $\sim30\%$ smaller than the model $h_{zn}$. This thicker exponential component of the H$\alpha$ profile corresponds with our measured $h_{z\mathrm{eDIG}}$, so following this we used $h_{zn} = 1.3 h_{z\mathrm{eDIG}}$ for our models. An example of the resulting density profile is shown for a model run of ESO~544-27 in Fig. \ref{fig:nH} alongside an HOLMES number-density profile.

\begin{figure*}[ht]
  \centering
  \includegraphics[width=\textwidth]{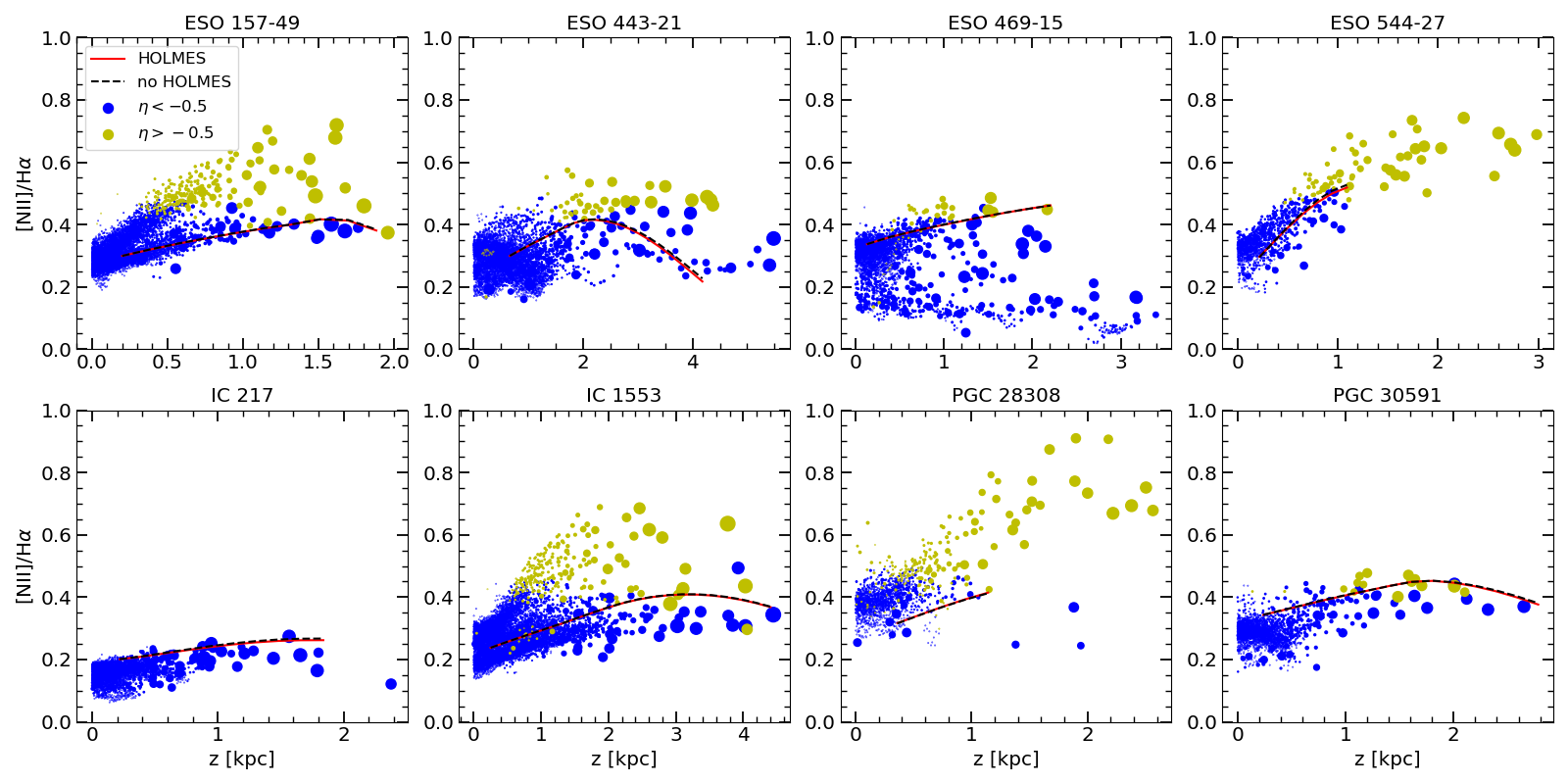}
  \caption{[N\,\textsc{ii}]/H$\alpha$ vertical profiles for our sample galaxies. The blue dots show observed values from Voronoi bins with OB star dominated ionization ($\eta < -0.5$). The yellow dots show observed values from Voronoi bins with mixed ionization or ionization dominated by other sources ($\eta > -0.5$). The size of each dot is relative to the size of the corresponding Voronoi bin. The solid red curve shows our model for the galaxy in question. The dashed black curve is the same model as the red curve but with HOLMES turned off. The contribution from HOLMES to the ionizing radiation is insignificant for our models when fitting OB star dominated ionization.}
  \label{fig:NIIvz}
\end{figure*}

\textsc{Starburst99} gives the total number of hydrogen ionizing photons emitted each second by a population with a certain SFR. With our parameters, it gave $10^{53.12}$ s$^{-1}$ for SFR of 1 M$_\odot$ yr$^{-1}$. By multiplying this number with the SFR of each galaxy, we obtained the total number of hydrogen ionizing photons emitted by that galaxy. Nearly all of these photons are produced by the midplane OB stars, with a fraction equal to $e_{\textrm{mp}}$ escaping the surrounding H{\sc ii} regions and reaching the eDIG. Dividing this number by $2\pi R^2$ gives the surface flux of hydrogen ionizing photons of OB stars at midplane ($\Phi_\mathrm{OB}$), assuming that the OB stars are evenly distributed in the H$\alpha$ disk and that half of the escaping photons are directed above the midplane and half of them below the midplane. Inputting $\Phi_\mathrm{OB}$ and the \textsc{Starburst99} spectrum into the preliminary H\,\textsc{ii} region \textsc{Cloudy} runs described in Sect. \ref{sec:ob} yielded $S_{\mathrm{OB}}(\lambda)$.

To obtain the total number of hydrogen ionizing photons emitted each second by HOLMES, we integrated the population-mass-normalized HOLMES spectrum given by \textsc{P\'egase}.3 and multiplied it with $M_\mathrm{T}$ of each galaxy. We then divided this number by $2\pi R^2$ to obtain the total $\Phi_\mathrm{HOLMES}$ produced by HOLMES on one side of the midplane, and distributed this between the cloud layers as described in Sect. \ref{sec:holmes} obtaining $S_{\mathrm{HOLMES},j}(\lambda)$.

{For our initial set of models we selected the values for our free parameters $U_1$ and $f_{V}$ by hand.} We did this by testing different values of $U_1$ and $f_{V}$ and comparing by eye the predicted vertical profiles of [N{\sc ii}]/H$\alpha$, [S{\sc ii}]/H$\alpha$, and [O{\sc iii}]/H$\beta$ to the observed line ratio profiles until we obtained a satisfactory match between them. We examine {these models} in detail in Sect. \ref{sec:results}. We did not attempt more rigorous numerical fitting {for these initial models} as our goal was to investigate the general validity of the proposed schema of ionization of the eDIG by leaking OB star radiation and HOLMES, rather than finding exact physical parameters of the eDIG. {We performed $\chi^2$ minimization on a grid of more detailed models for two of the galaxies as described in Sect.~\ref{sec:hcon}.}

\section{Results}
\label{sec:results}

\subsection{Ionization of the eDIG in the sample galaxies}

\begin{figure*}[ht]
  \centering
  \includegraphics[width=\textwidth]{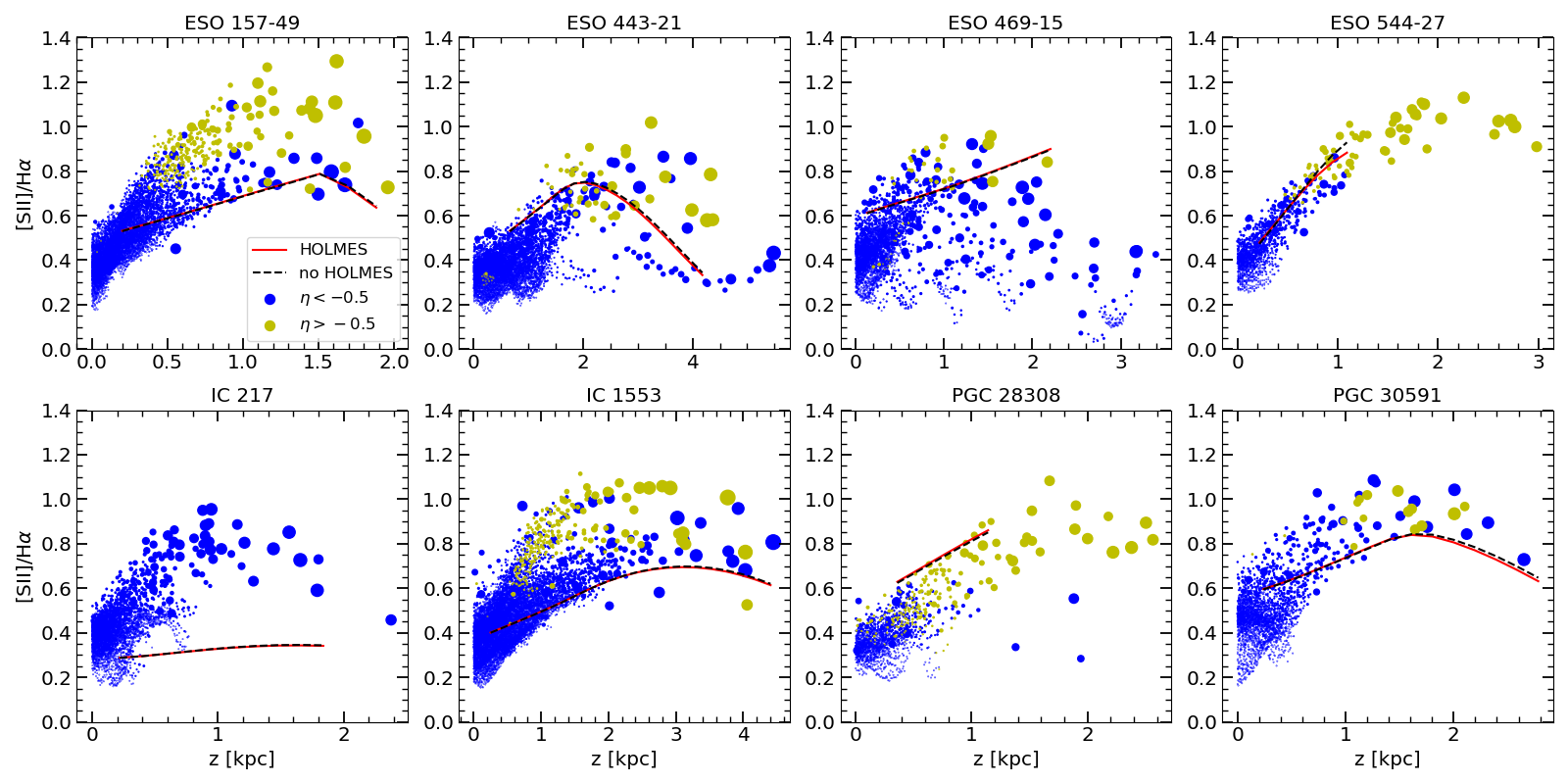}
  \caption{Same as Fig. \ref{fig:NIIvz} but for [S\,\textsc{ii}]/H$\alpha$.}
  \label{fig:SIIvz}
\end{figure*}

\begin{figure*}[ht]
  \centering
  \includegraphics[width=\textwidth]{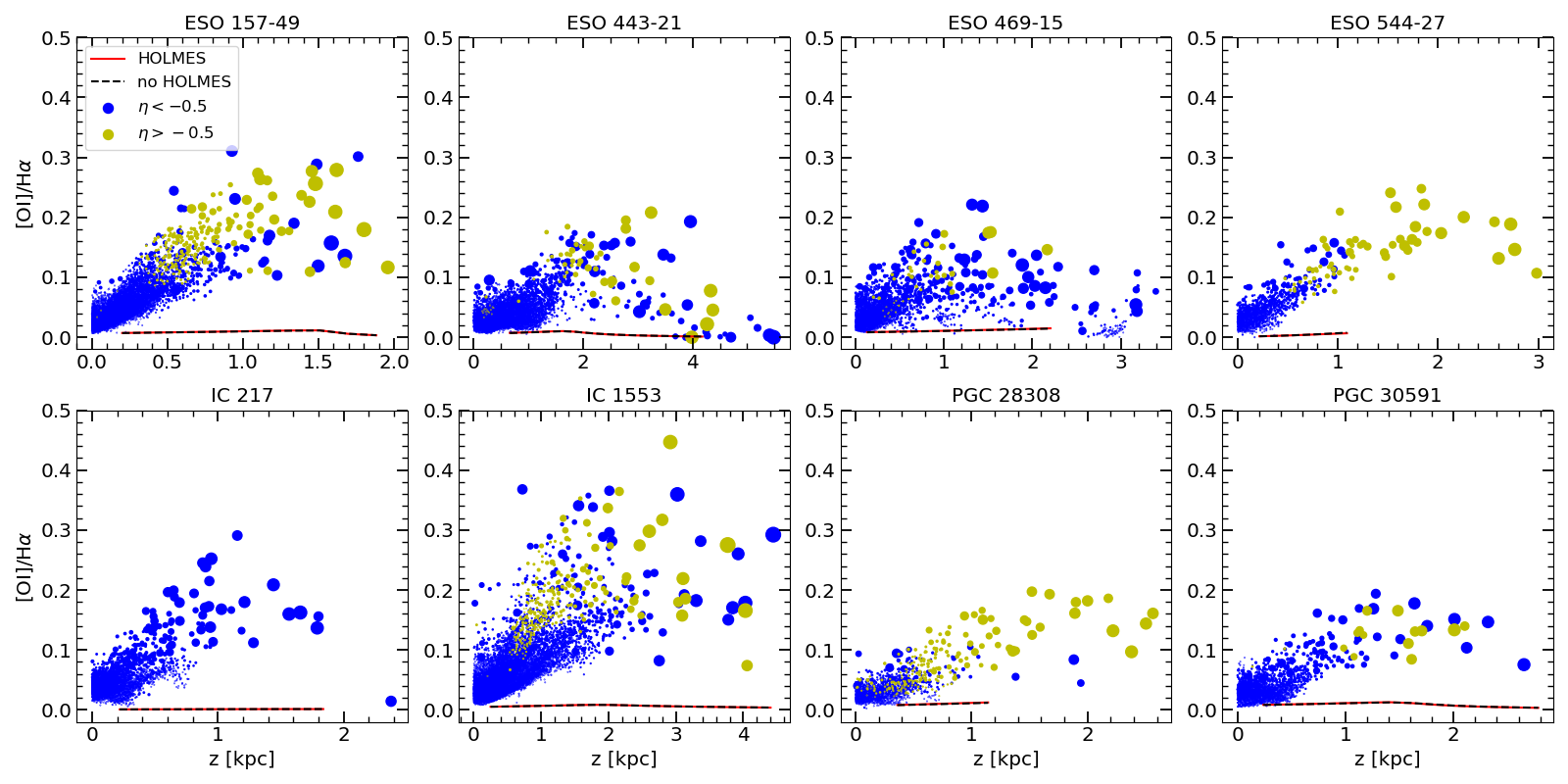}
  \caption{Same as Fig. \ref{fig:NIIvz} but for [O\,\textsc{i}]/H$\alpha$. The [O\,\textsc{i}]/H$\alpha$ line ratio is poorly fit because [O\,\textsc{i}] is a collisionally excited line of neutral oxygen, and as such is only produced by partly ionized and neutral gas, while our model emission lines are dominated by fully ionized gas.}
  \label{fig:OIvz}
\end{figure*}

\begin{figure*}[ht]
  \centering
  \includegraphics[width=\textwidth]{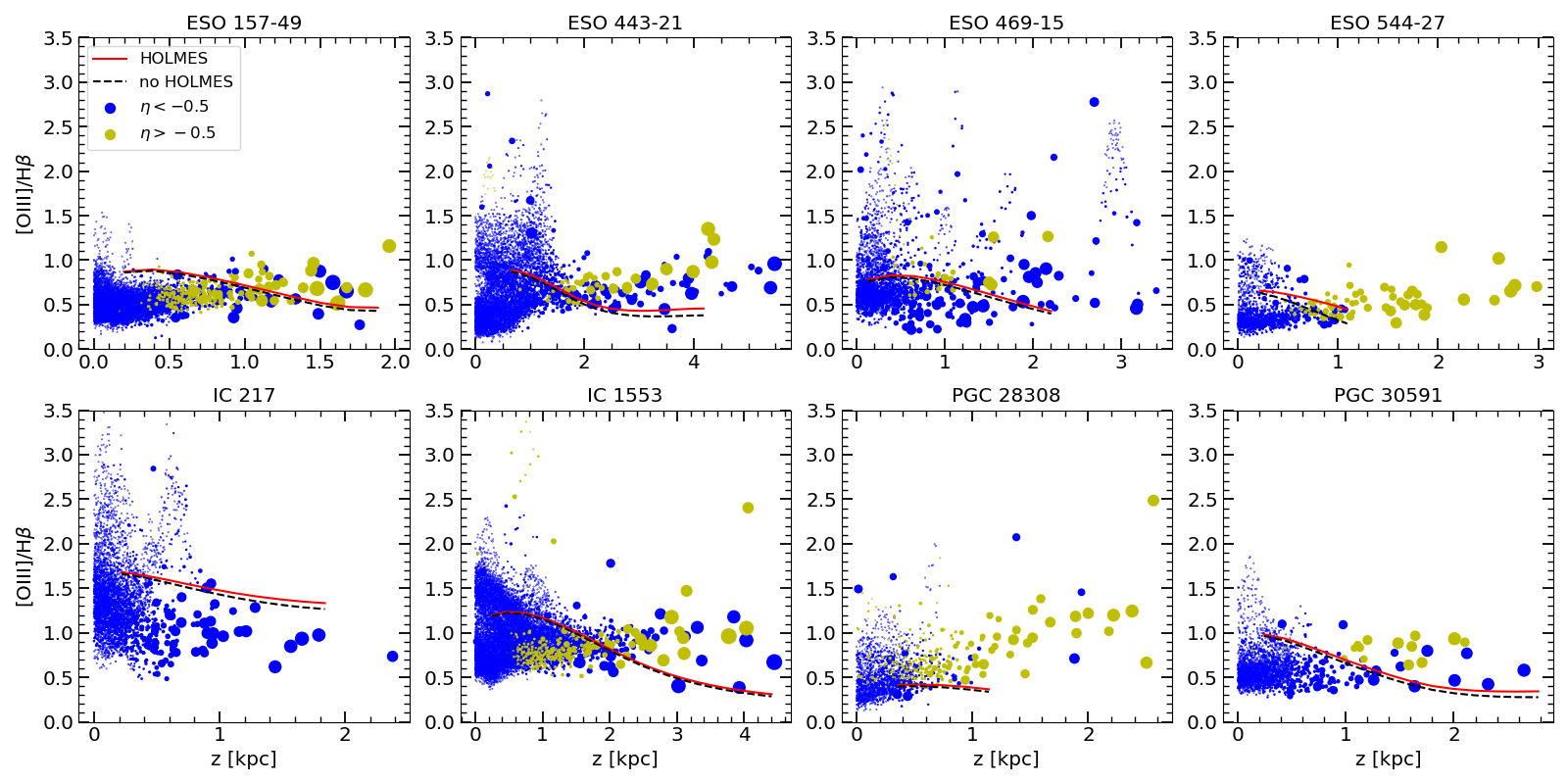}
  \caption{Same as Fig. \ref{fig:NIIvz} but for [O\,\textsc{iii}]/H$\beta$.}
  \label{fig:OIIIvz}
\end{figure*}

Figures \ref{fig:NIIvz}, \ref{fig:SIIvz}, \ref{fig:OIvz}, and \ref{fig:OIIIvz}. show the vertical profiles of the [N{\sc ii}]/H$\alpha$, [S{\sc ii}]/H$\alpha$, [O{\sc i}]/H$\alpha$, and [O{\sc iii}]/H$\beta$ line ratios for our sample galaxies. The blue and yellow dots are the observed values from MUSE data, with each dot corresponding to a single Voronoi bin, and the size of the dot corresponding with the size of the Voronoi bin. The blue dots are from Voronoi bins where the ionization is dominated by OB star radiation, while the yellow dots are from bins with mixed ionization or ionization dominated by other sources. We did this separation using the $\eta$-parameter, so that blue dots have $\eta < -0.5$ and yellow dots have $\eta > -0.5$. {The red curves correspond to our model runs.} The black curves are the same as the red curves, only with HOLMES turned off. {When choosing the free parameters}, we only considered the bins with OB star dominated ionization ($\eta < -0.5$, blue dots). We did this because some  of the galaxies in our sample exhibit clear shock ionization \citepalias{rautio2022}, and here we are only modeling photoionization. In particular ESO~157-49 and IC~1553 exhibit clear bimodal distributions in their vertical [N{\sc ii}]/H$\alpha$ profiles, and the branches with enhanced $\eta$ correspond spatially with biconical outflows visible in the $\eta$-parameter and velocity dispersion maps. These branches with enhanced $\eta$ are most likely caused by shock ionization as a superbubble breaks out from the midplane (for {IC~1553} see \citealt{dirks2023sb}), and as such we ignored them when fitting our photoionization model to ESO~157-49 and IC~1553.

\begin{figure*}[ht]
  \centering
  \includegraphics[width=\textwidth]{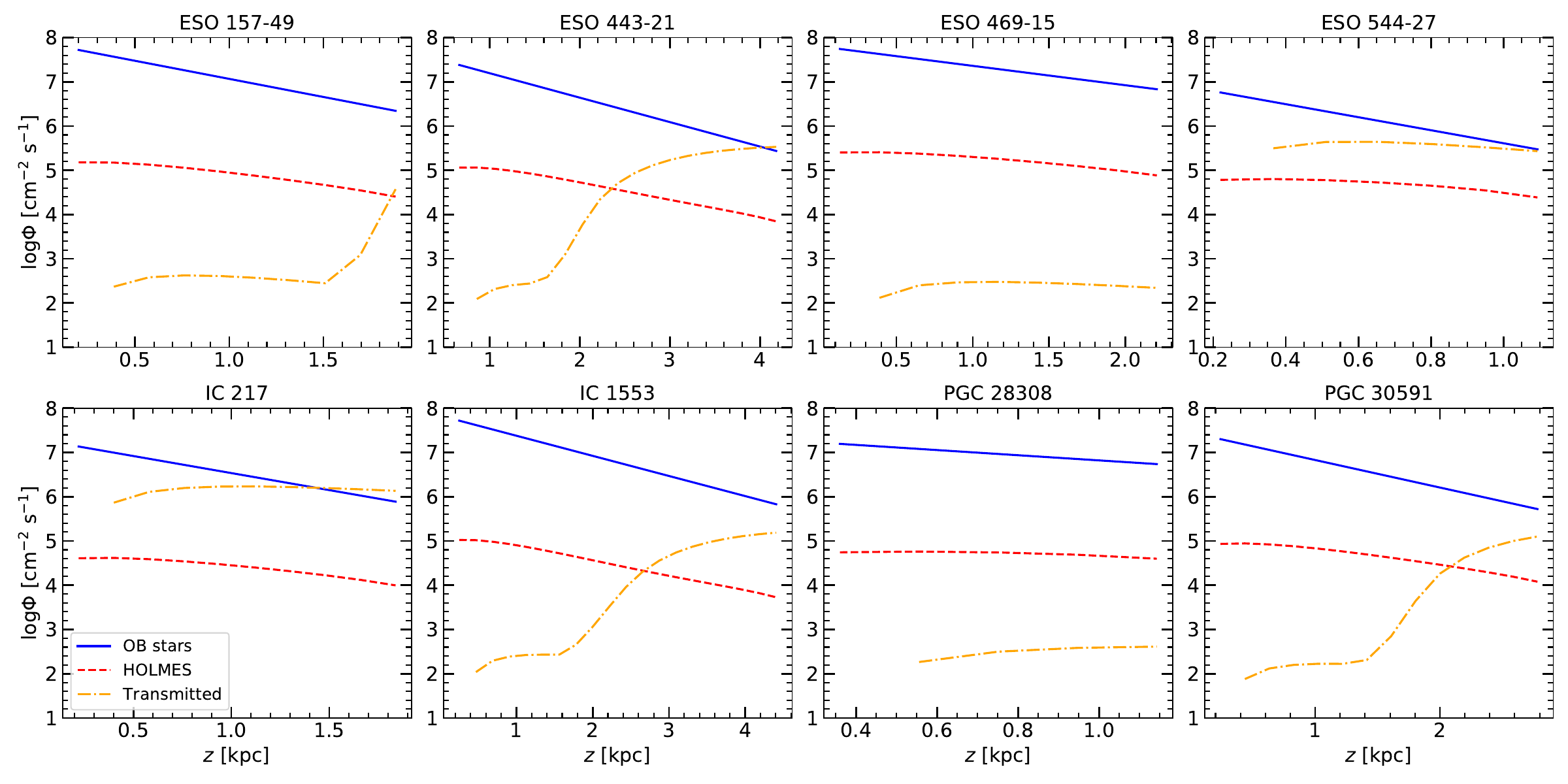}
  \caption{Contributions of different components to the ionizing radiation for the best fit models. Blue solid lines shows the contribution from midplane OB stars, red dashed line shows the contribution from HOLMES, and the orange dot-dashed line shows the contribution of transmitted radiation. The transition from ionization bounded clouds to density bounded clouds is visible as a steep upturn in the contribution of transmitted radiation.}
  \label{fig:phiall}
\end{figure*}

Figure \ref{fig:phiall} shows the contribution of different ionization sources to the ionizing radiation in the models. The transmitted component curve in the figure (orange dot-dashed line) includes both the transmitted OB star radiation and the transmitted HOLMES radiation. For all of the galaxies $\Phi_{\mathrm{HOLMES}} \le 10^{-2}\Phi_{\mathrm{OB}}$ for the first cloud, with the contribution of HOLMES increasing slowly with $z$ and reaching $\Phi_{\mathrm{HOLMES}} \approx 10^{-1}\Phi_{\mathrm{OB}}$ at most (for ESO~544-27). The contribution of transmitted radiation depends on whether the eDIG clouds are fully ionized and density bounded or ionization bounded. For ESO~469-15 and PGC~28308 models all of the eDIG clouds are ionization bounded, and very little radiation is transmitted, resulting in very low $\Phi_{\mathrm{t}}$ at all $z$. For ESO~544-27 and IC~217 models all of the eDIG clouds are density bounded, and significant portion of the radiation is transmitted at every layer. For ESO~157-49, ESO~443-21, IC~1553, and PGC~30591 eDIG clouds near midplane are ionization bounded, while as $z$ increases and density goes down, upper eDIG clouds become density bounded, resulting in a clear transition in the $\Phi_{\mathrm{t}}$ profile where it increases significantly when density becomes low enough to allow for the full ionization of the eDIG clouds.

\begin{figure}
  \centering
  \includegraphics[width=0.5\textwidth]{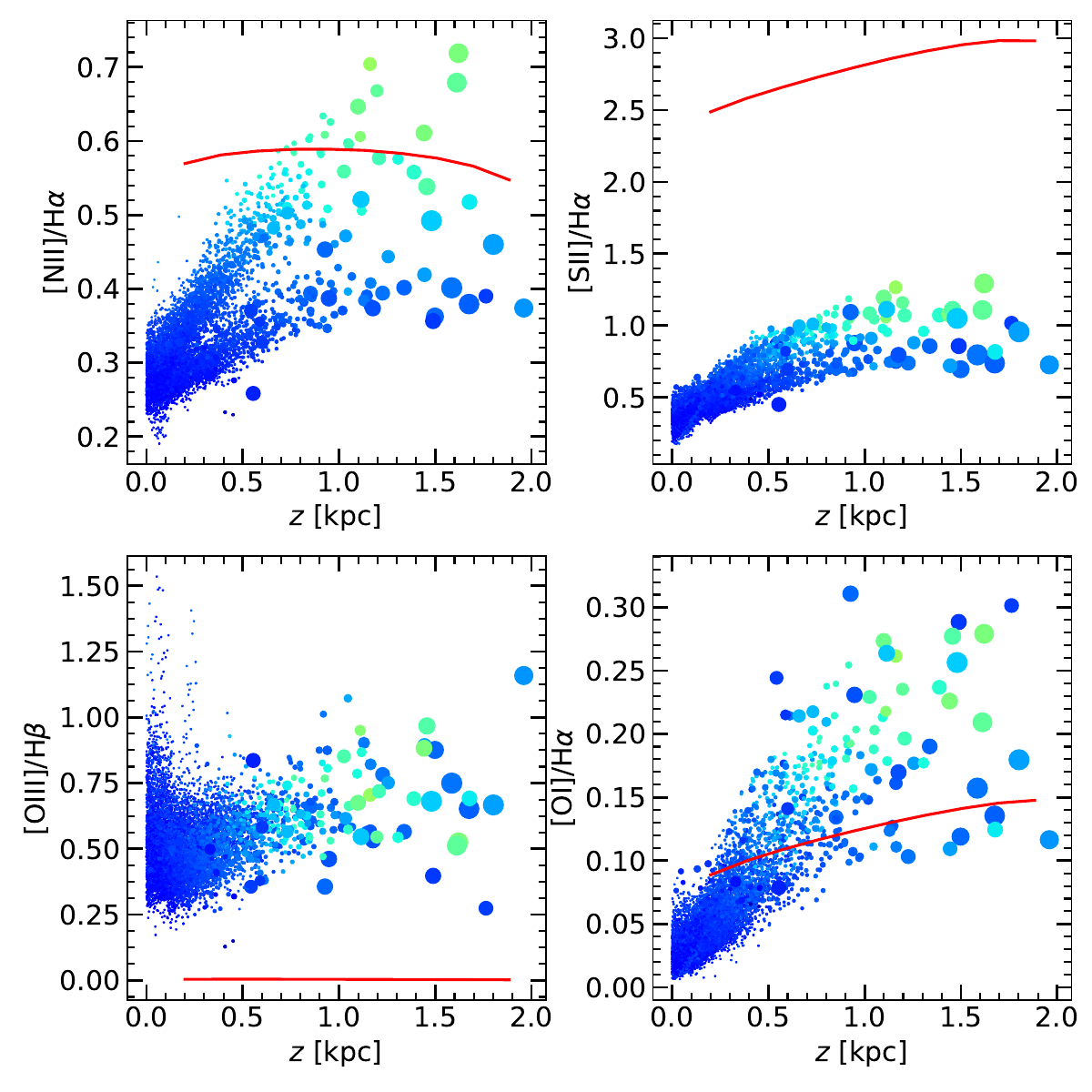}
  \caption{Line ratio profiles for ESO~157-49 model with 2\% $\Phi_{\mathrm{OB}}$. The dots show the observed values, colored according to the $\eta$-parameter going from blue at low $\eta$ to green at high $\eta$. The size of each dot is relative to the size of the corresponding Voronoi bin. The red curve shows the model predictions. This model predicts the [O\,\textsc{i}]/H$\alpha$ line ratio well, but fails to predict any of the other line ratios. The $\log U_1 = -4.51$ and other parameters are same as in the best fit model.}
  \label{fig:oi}
\end{figure}

The models were chosen to best {reproduce} [N{\sc ii}]/H$\alpha$, [S{\sc ii}]/H$\alpha$, and [O{\sc iii}]/H$\beta$ simultaneously, ignoring [O{\sc i}]/H$\alpha$, as no model was able to simultaneously reproduce all of the observed line ratios. This is because [O{\sc i}] is a collisionally excited line of neutral oxygen, and as such is only produced by partly ionized and neutral gas. Our \textsc{Cloudy} models do not extend far into neutral gas, and as such even in the models with only ionization bounded eDIG clouds the majority of the emitted radiation is produced in the fully ionized parts of the clouds. In the real complex 3D environment of the eDIG there exist simultaneously fully ionized, partly ionized, and neutral gas. Thus, as we are fitting the line ratios originating from fully ionized gas, our models underestimate [O{\sc i}]/H$\alpha$. As an example of a model that {reproduces} [O{\sc i}]/H$\alpha$ well, Fig. \ref{fig:oi} shows the line ratio profiles for an ESO~157-49 model where $\Phi_{\mathrm{OB}}$ has been reduced to 2\% of our prediction using SFR and \textsc{Starburst99} modeling. In this model the transition region to neutral gas, where [O{\sc i}] is produced, is of comparable size to the fully ionized part of the cloud. For this model we kept the density of the gas equal to the best fit model by setting $\log U_1 = -4.51$. Other parameters of this model are the same as the best fit model.

We were unable to {simultaneously reproduce} both of the observed [N{\sc ii}]/H$\alpha$ and [S{\sc ii}]/H$\alpha$ for {some} of the galaxies in our sample, particularly IC~217 and PGC~28308 (see Fig. \ref{fig:NIIvz} and Fig. \ref{fig:SIIvz}). This is most likely because of the simplified approach to the elemental abundances in our models. We are using the preset interstellar medium ``ISM'' abundances in \textsc{Cloudy}, and only scaling the elements heavier than helium to match the observed metallicity ($12+\log(\mathrm{O}/\mathrm{H}$)). This does not take into account any differences in nitrogen and sulfur abundances, which are important to accurately model the [N{\sc ii}] to [S{\sc ii}] ratio. This is not an issue for [O{\sc iii}], as the second ionization potential of oxygen is significantly higher than the first ionization potentials of [N{\sc ii}] and [S{\sc ii}] (35.1 eV, 14.5 eV, and 10.4 eV, respectively), and as such the ratio of [O{\sc iii}] to [N{\sc ii}] or [S{\sc ii}] is less dependent on elemental abundances and more on the shape and intensity of the ionizing spectrum.

\begin{table}
  \caption{{Model parameters.}}
  \label{tab:params}
  \centering
  \begin{tabular}{lccccccc}
    \hline\hline
    ID & $\log{U_1}$ & $f_V$ & $h_{zn}$ & $R$ & $N$ \\

    & & & (kpc) & (kpc) & \\ 
    \hline
    ESO~157-49 & -2.81 & 0.08 & 0.78 & 2.0 & 10\\ 
    ESO~443-21 & -2.81 & 0.07 & 1.40 & 10.5 & 19\\ 
    ESO~469-15 & -2.90 & 0.03 & 1.72 & 2.0 & 9\\ 
    ESO~544-27 & -3.01 & 0.17 & 0.79 & 5.0 & 7 \\ 
    IC~217 & -2.86 & 0.09 & 1.46 & 4.5 & 10 \\ 
    IC~1553 &  -2.64 & 0.05 & 1.81 & 4.5 & 20 \\ 
    PGC~28308 & -2.95 & 0.07 & 1.40 & 6.0 & 5 \\ 
    PGC~30591 & -2.91 & 0.07 & 1.18 & 4.0 & 14 \\ 
    \hline
  \end{tabular}
  \tablefoot{$U_1$ is the ionization parameter at the surface of the first cloud, $f_V$ is the volume filling factor, $h_{zn}$ is the scale height of the gas density, $R$ is the radius of the galaxy, and $N$ is the number of cloud layers.}
\end{table}

Table \ref{tab:params} shows the {chosen} parameters for each galaxy. The gas densities of the first cloud layers in our models range from 0.2 cm$^{-3}$ to 1 cm$^{-3}$. These are somewhat high when compared to the measured DIG densities in Milky Way ($0.34\pm0.06$ cm$^{-3}$ at midplane; \citealt{gaensler2008wim}). However, for all models the gas density falls below 0.3 cm$^{-3}$ higher in the eDIG, where our simple 1D modeling scenario works best. The chosen $f_V$ for our galaxies on the other hand are lower than the measured $f_V$ for Milky Way ($\sim0.2$; e.g., \citealt{reynolds1991ff}). Similar results are found in studies of face-on galaxies, that extremely low $f_V$ are required to explain the propagation of OB star radiation and ionization of DIG within the midplane \citep{seon2009dig, belfiore2022dig, watkins2024m101}.

\subsection{Contribution of HOLMES to the eDIG ionization}
\label{sec:hcon}

Figures \ref{fig:NIIvz}, \ref{fig:SIIvz}, \ref{fig:OIvz}, and \ref{fig:OIIIvz} show very little difference between the models with and without HOLMES. This is to be expected when modeling OB star dominated ionization. In \citetalias{rautio2022} we suggest that HOLMES could be responsible for the enhanced $\eta$-parameter in the eDIG in ESO~544-27. ESO~544-27 has a relatively massive thick disk and relatively low SFR when compared to the other galaxies in our sample, meaning that the HOLMES contribution to its ionization budget is much larger, especially in the eDIG. To investigate this we constructed an alternative high-$\Phi_{\mathrm{HOLMES}}$ model for ESO~544-27, this time attempting to fit the {Voronoi bins with mixed ionization source ($\eta > -0.5$).} For this model we set the HOLMES mass to $1.5\times M_\mathrm{T}$, $\Phi_{\mathrm{OB}}$ to half of what our \textsc{Starburst99} models predict, and $e_{\mathrm{mp}} = 0.2$. These alternate parameters result in much higher $\Phi_{\mathrm{HOLMES}}$ compared to $\Phi_{\mathrm{OB}}$, while still being reasonable considering the large uncertainties in determining the photon fluxes of our models. We also left $h_{zn}$ as a free parameter, as we found that it is the only parameter that can move the peak of the [N{\sc ii}]/H$\alpha$ and [S{\sc ii}]/H$\alpha$ line ratio profiles to higher $z$.

Since the HOLMES contribution to the transmitted radiation field is much higher for the high-$\Phi_{\mathrm{HOLMES}}$ ESO~544-27 model than to any of the previous models, we adopted a more rigorous treatment for the transmitted radiation field. To do this we followed an iterative approach, where first for each cloud layer we divided the output transmitted radiation field to two components with intensity ratio equal to the input $\Phi_{\mathrm{HOLMES}}/\Phi_{\mathrm{OB}}$, then we further divided the component related to $\Phi_{\mathrm{HOLMES}}$ into upwards component and downwards component, which were proportional to the amount of HOLMES below the cloud layer and above the cloud layer, respectively. We then used the downward HOLMES component of the transmitted radiation field as an additional input in a re-run of the model. This essentially modified Eq. \ref{eq:trans} for the re-run into

\begin{equation}
  \begin{aligned}
    I_{\mathrm{t},i}^1(\lambda) = & \sum_{j < i} g_{ij} (1-f_S)^{|j-i|} \left(1-a_j\frac{\Phi_{\mathrm{HOLMES},j}}{\Phi_{\mathrm{OB},j}}\right) f_S S_{\mathrm{t},j}^1(\lambda) \\
    & + \sum_{j > i} g_{ij} (1-f_S)^{|j-i|} a_j \frac{\Phi_{\mathrm{HOLMES},j}}{\Phi_{\mathrm{OB},j}} f_S S_{\mathrm{t},j}^0(\lambda), \\
    & i = 1,\dotsc,N,
  \end{aligned}
\end{equation}

\noindent where $a_j$ is the fraction of HOLMES above layer $j$, $S_{\mathrm{t},j}^0(\lambda)$ is the source transmitted radiation field from the first run, and $I_{\mathrm{t},i}^1(\lambda)$ and $S_{\mathrm{t},j}^1(\lambda)$ are the incident and source transmitted radiation fields of the second run. We tested further iterations and found that the model converges quickly and the changes to the line ratio profiles are indistinguishable by eye after the first iteration.

{We constructed a 1100 model grid of high-$\Phi_{\mathrm{HOLMES}}$ ESO~544-27 models by varying $h_{zn}$ from 0.8 kpc to 1.8 kpc in 0.1 kpc steps, $\log U_1$ from -3.35 to -2.90 in steps of 0.5, and $f_V$ from 0.05 to 0.32 in steps of 0.03. We found the best fit model by minimizing the $\chi^2$ defined as}

\begin{equation}
  \chi^2 = \sum^{3n} \frac{[I_\mathrm{1, obs}/I_\mathrm{2, obs} - I_\mathrm{1, model}/I_\mathrm{2, model}]^2}{\sigma_{1,2}^2},
\end{equation}

\noindent {where $n$ is the number of Voronoi bins with $\eta > -0.5$, $I_\mathrm{1, obs}/I_\mathrm{2, obs}$ are the observed [N{\sc ii}]/H$\alpha$, [S{\sc ii}]/H$\alpha$, and [O{\sc iii}]/H$\beta$ line ratios of those Voronoi bins, $I_{1, \mathrm{model}}/I_{2, \mathrm{model}}$ are the model values for those line ratios, and $\sigma_{1,2}^2$ are the uncertainties of the line ratios, defined as}

\begin{equation}
    \sigma_{1,2}^2 =  \left(\frac{I_\mathrm{1, obs}}{I_\mathrm{2, obs}}\right)^2 \left[\left(\frac{\sigma_1}{I_\mathrm{1, obs}}\right)^2 + \left(\frac{\sigma_2}{I_\mathrm{2, obs}}\right)^2\right].
\end{equation}

\noindent {We restricted the fitting to Voronoi bins with $\eta > -0.5$ because we are interested in finding a model that reproduces the parts of eDIG with mixed ionization sources. Without this restriction $\chi^2$ would be heavily weighed by values near the midplane due to the better signal-to-noise and larger number of Voronoi bins therein.}

\begin{figure}
  \centering
  \includegraphics[width=0.5\textwidth]{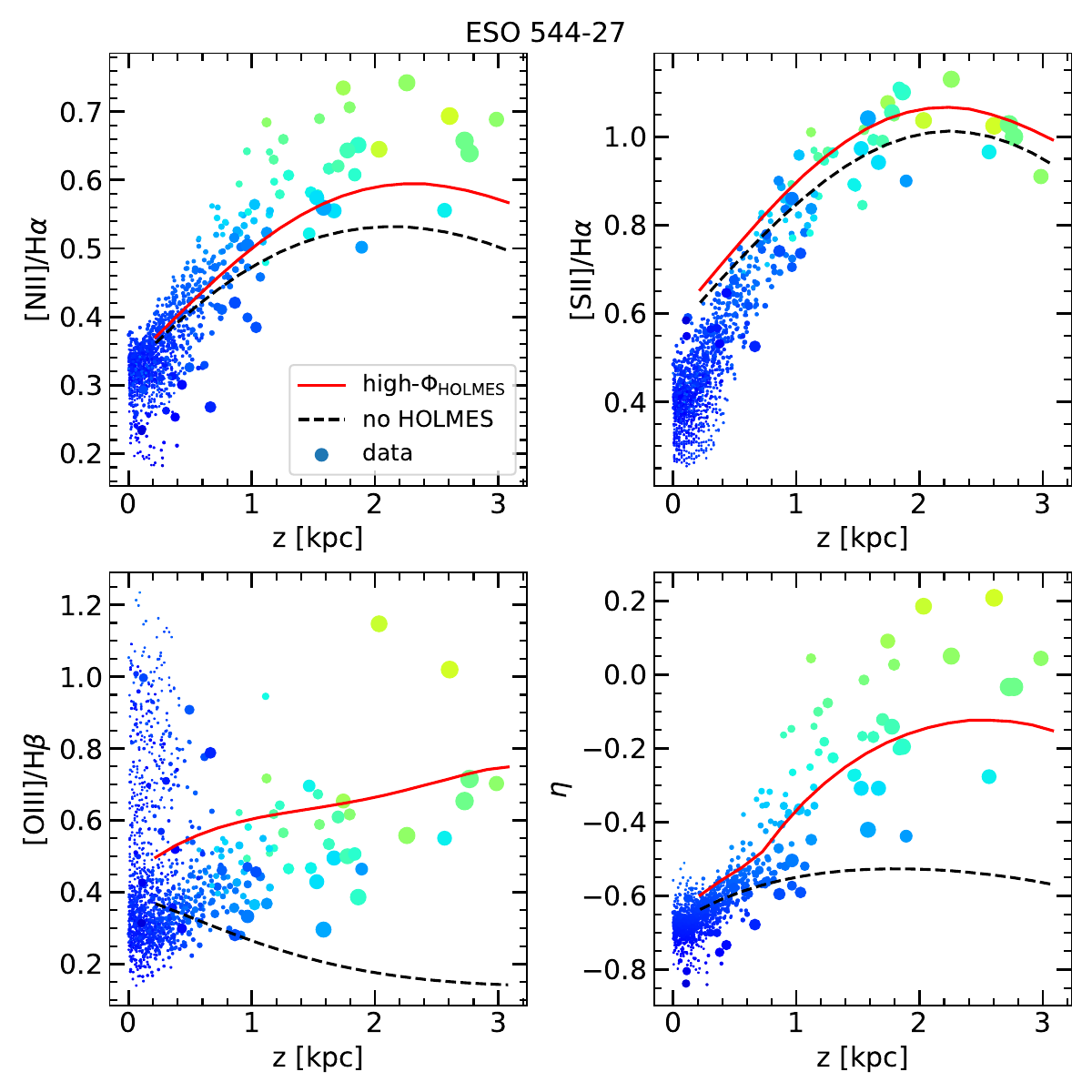}
  \caption{{Line ratio and $\eta$-parameter profiles for the best fit alternative high-$\Phi_{\mathrm{HOLMES}}$ ESO~544-27 model (see text). The dots show the observed values, colored according to the $\eta$-parameter going from blue at low $\eta$ to green at high $\eta$. The size of each dot is relative to the size of the corresponding Voronoi bin. The solid red curve shows the best fit model including HOLMES ($\chi^2 / \nu = 39$) while the dashed black curve shows the best fit model without HOLMES ($\chi^2 / \nu = 75$). The best fit values for the models including HOLMES are $h_{zn} = 1.7$ kpc, $f_V = 0.11$, and $\log U_1 = -3.30$, while the best fit values for the models without HOLMES are $h_{zn} = 1.6$ kpc, $f_V = 0.11$, and $\log U_1 = -3.30$. The high-$\Phi_{\mathrm{HOLMES}}$ model reproduces the observed $\eta > -0.5$ in the eDIG of ESO~544-27.}}
  \label{fig:2x}
\end{figure}

{We found the best fit with $h_{zn} = 1.7$ kpc, $f_V = 0.11$, and $\log U_1 = -3.30$. The goodness of the fit is $\chi^2 / \nu = 39$, which is normalized by the number of degrees of freedom\footnote{$\nu = 3n - k$ is the number of degrees of freedom, where $k$ is the number of free parameters}. In order to verify that HOLMES are required to find a good fit, we also constructed a 1100 model grid of ESO~544-27 models with otherwise the same parameters, but HOLMES turned off. For this grid we found the best fit with $h_{zn} = 1.6$ kpc, $f_V = 0.11$, and $\log U_1 = -3.30$, with $\chi^2 / \nu = 75$. The best fit models from both of these grids are shown in Fig. \ref{fig:2x}. As $\chi^2 / \nu$ is nearly twice as large for the best fit model without HOLMES compared to the high-$\Phi_{\mathrm{HOLMES}}$ best fit model, and it is visually clear from Fig. \ref{fig:2x} that the high-$\Phi_{\mathrm{HOLMES}}$ best fit model reproduces $\eta$ and the line ratios much better than the best fit model without HOLMES, we confirm that including HOLMES improves the fit for ESO~544-27.}

\begin{figure}
  \centering
  \includegraphics[width=0.5\textwidth]{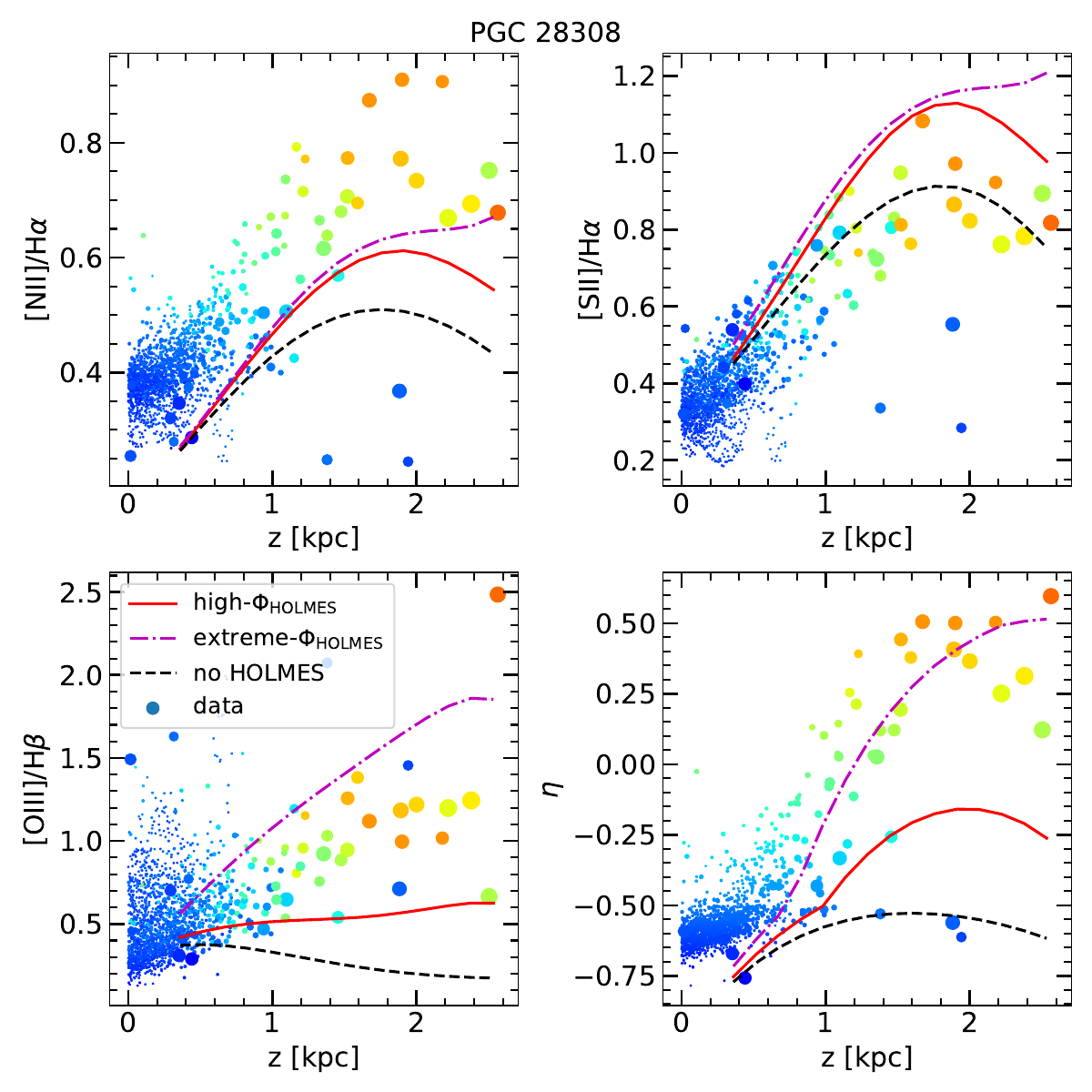}
  \caption{{Line ratio and $\eta$-parameter profiles for the best fit PGC~28308 models. The dots show the observed values, colored according to the $\eta$-parameter going from blue at low $\eta$ to green at high $\eta$. The size of each dot is relative to the size of the corresponding Voronoi bin. The solid red curve shows the best fit high-$\Phi_{\mathrm{HOLMES}}$ model ($\chi^2 / \nu = 156$), the dash-dot magenta curve shows the best fit extreme-$\Phi_{\mathrm{HOLMES}}$ model ($\chi^2 / \nu = 167$), and the dashed black curve shows the best fit model without HOLMES ($\chi^2 / \nu = 175$). The best fit values are $h_{zn} = 1.2$ kpc, $f_V = 0.14$, and $\log U_1 = -3.10$ for the high-$\Phi_{\mathrm{HOLMES}}$ model, $h_{zn} = 1.9$ kpc, $f_V = 0.14$, and $\log U_1 = -3.20$ for the extreme-$\Phi_\mathrm{HOLMES}$ model, and $h_{zn} = 1.1$ kpc, $f_V = 0.14$, and $\log U_1 = -3.10$ for the model without HOLMES. The reasonable high-$\Phi_{\mathrm{HOLMES}}$ model fails to reproduce the high $\eta$-parameter in the eDIG of PGC~28308, and a model with unrealistically high $\Phi_\mathrm{HOLMES}/\Phi_\mathrm{OB}$ is required to do so. }}
  \label{fig:2x2}
\end{figure}

{Another galaxy in our sample with a similar enhancement of the $\eta$-parameter throughout the eDIG as ESO~544-27 is PGC~28308. We constructed a similar high-$\Phi_{\mathrm{HOLMES}}$ and no HOLMES model grids for it by varying $h_{zn}$ from 1.0 kpc to 2.0 kpc in 0.1 kpc steps, $\log U_1$ from -3.35 to -2.90 in steps of 0.5, and $f_V$ from 0.05 to 0.32 in steps of 0.03. The high-$\Phi_{\mathrm{HOLMES}}$ models also had $1.5\times M_\mathrm{T}$ HOLMES mass, halved $\Phi_{\mathrm{OB}}$, $e_{\mathrm{mp}} = 0.2$, and used the more rigorous treatment of the transmitted radiation as described above. By minimizing $\chi^2$ we again found the best fit high-$\Phi_{\mathrm{HOLMES}}$ model for $\eta > -0.5$ with $h_{zn} = 1.2$ kpc, $f_V = 0.14$, and $\log U_1 = -3.10$, and best fit model without HOLMES for $\eta > -0.5$ with $h_{zn} = 1.1$ kpc, $f_V = 0.14$, and $\log U_1 = -3.10$. However, for this high-$\Phi_{\mathrm{HOLMES}}$ fit $\chi^2 / \nu = 156$, which while higher than the best fit no HOLMES model ($\chi^2 / \nu = 175$), is much worse than $\chi^2$ of the ESO~544-27 best fit models, and the model fails to reproduce the high $\eta$-parameter in the eDIG of PGC~28308. In order to find a better fit we constructed an extreme-$\Phi_{\mathrm{HOLMES}}$ PGC~28308 model grid with otherwise the same parameters as the high-$\Phi_{\mathrm{HOLMES}}$ PGC~28308 model grid except with increased HOLMES mass to $4\times M_\mathrm{T}$ and quartered $\Phi_{\mathrm{OB}}$. We found for the best fit extreme-$\Phi_{\mathrm{HOLMES}}$ PGC~28308 model $h_{zn} = 1.9$ kpc, $f_V = 0.14$, and $\log U_1 = -3.20$, and while this model is able to reproduce the observed high $\eta$-parameter in the eDIG of PGC~28308, it is still not very well fit with $\chi^2 / \nu = 167$. The best fit models for PGC~28308 are shown in Fig. \ref{fig:2x2}.}

\begin{figure*}
  \centering
  \includegraphics[width=\textwidth]{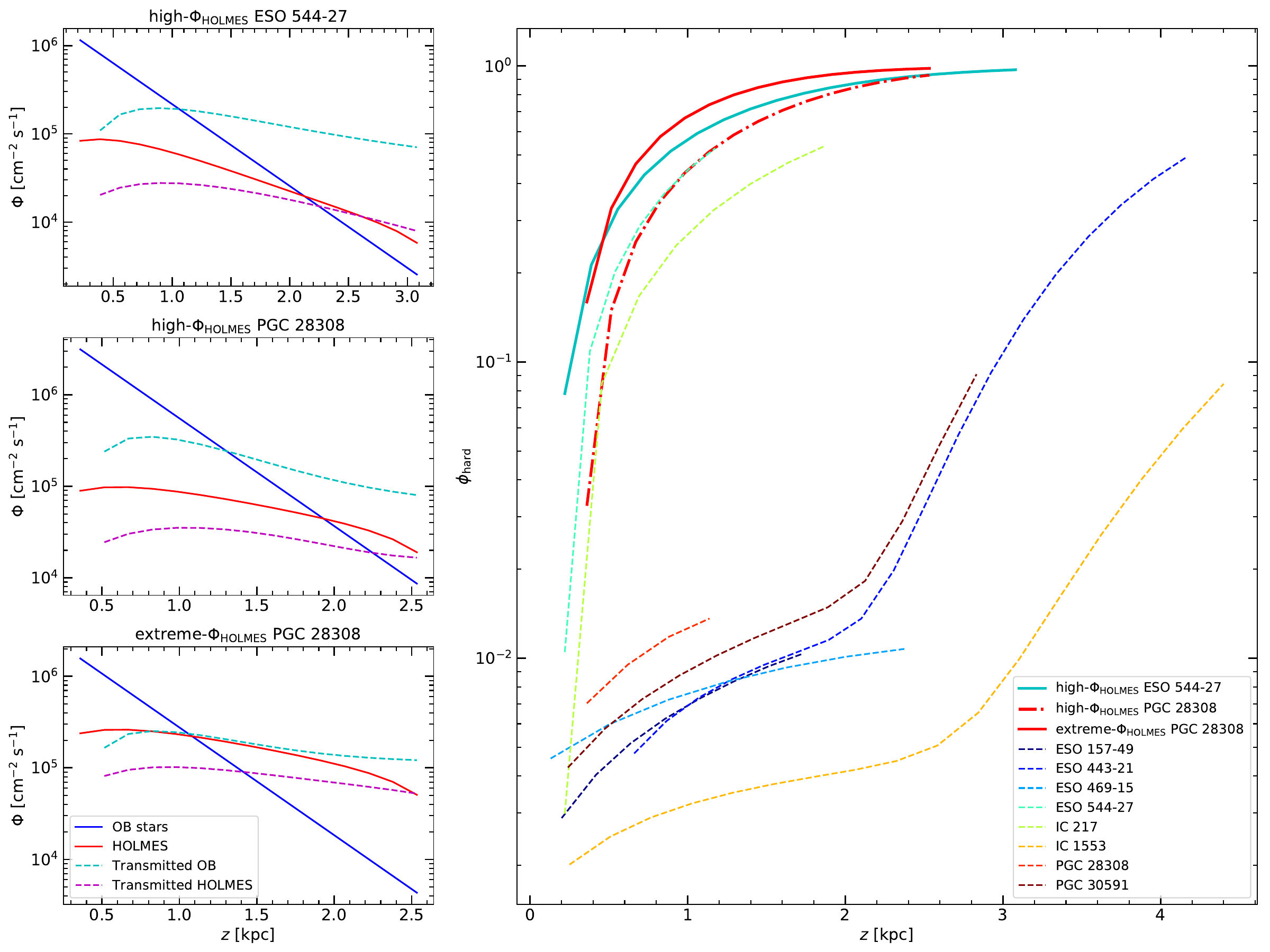}
  \caption{Contributions of the different ionization sources over $z$ for the models with enhanced $\Phi_\mathrm{HOLMES}$. \emph{Top-left panel:} contributions of different components to the ionizing radiation for the high-$\Phi_{\mathrm{HOLMES}}$ ESO~544-27 model. The blue solid line shows the contribution from midplane OB stars, the red solid line shows the contribution from HOLMES, and the cyan and magenta dashed lines show the contribution of transmitted emission from midplane OB stars and HOLMES, respectively. \emph{Middle-left panel:} same as top-left panel but for the high-$\Phi_{\mathrm{HOLMES}}$ PGC~28308 model. \emph{Bottom-left panel:}  same as top-left panel but for the extreme-$\Phi_\mathrm{HOLMES}$ PGC~28308 model. \emph{Right panel:} relative contribution of combined HOLMES radiation and transmitted radiation for all of our models $\phi_\mathrm{hard} = (\Phi_\mathrm{HOLMES}+\Phi_\mathrm{t})/\Phi$. The thick cyan curve shows $\phi_\mathrm{hard}$ for the high-$\Phi_{\mathrm{HOLMES}}$ ESO~544-27 model. The thick dash-dot red curve shows $\phi_\mathrm{hard}$ for the high-$\Phi_{\mathrm{HOLMES}}$ PGC~28308 model. The thick solid red curve shows $\phi_\mathrm{hard}$ for the extreme-$\Phi_\mathrm{HOLMES}$ PGC~28308 model. The dashed curves show $\phi_\mathrm{hard}$ for the original models. Only models where $\phi_\mathrm{hard}$ approaches unity are able to produce $\eta > -0.5$ in the eDIG.}
  \label{fig:phi}
\end{figure*}

In order to determine quantitatively how high HOLMES flux is required to produce the observed $\eta$-parameter in the eDIG of ESO~544-27 and PGC~28308 we looked into the relative contributions of different ionization sources in our models. These are plotted in Fig. \ref{fig:phi}. The left panels show the contributions of different components to the ionizing radiation in the eDIG of our high-$\Phi_{\mathrm{HOLMES}}$ models, with the transmitted radiation shown separately for OB stars and HOLMES. These plots show that for the models that are capable of reproducing the observed enhanced $\eta$-parameter in the eDIG of ESO~544-27 and PGC~28308 (high-$\Phi_{\mathrm{HOLMES}}$ ESO~544-27 model and the extreme-$\Phi_\mathrm{HOLMES}$ PGC~28308 model), {$\Phi_\mathrm{OB}$ is lower than $\Phi_\mathrm{HOLMES}$ or $\Phi_\mathrm{t}$ at high $z$. However, this is also true for the high-$\Phi_{\mathrm{HOLMES}}$ PGC~28308 model that cannot reproduce the observed enhanced $\eta$-parameter at high $z$. Also, for the high-$\Phi_{\mathrm{HOLMES}}$ ESO~544-27 model, $\Phi_\mathrm{t}$ is still an order of magnitude higher than $\Phi_{\mathrm{HOLMES}}$ at high $z$, meaning that most of the ionizing flux there still originates from the midplane. This would seem to suggest that both the radiation from in situ HOLMES and the procession of midplane radiation by intermediate clouds contribute to the hardening of the radiation field and enhancement of the $\eta$-parameter observed in eDIG.}

The right panel of Fig. \ref{fig:phi} shows the combined relative contribution of HOLMES and transmitted radiation to the ionizing radiation in the eDIG of all our models $\phi_\mathrm{hard} = (\Phi_\mathrm{HOLMES}+\Phi_\mathrm{t})/\Phi$. The combined relative HOLMES and transmitted radiation contribution approaches unity in the eDIG of the {high-$\Phi_{\mathrm{HOLMES}}$ ESO~544-27 model, and the high-$\Phi_{\mathrm{HOLMES}}$ and extreme-$\Phi_\mathrm{HOLMES}$ PGC~28308 models, while remaining below 0.6 for all other models.} This suggests that the contribution of unprocessed midplane radiation to the ionization of eDIG must be vanishingly low where $\eta > -0.5$.

\section{Discussion}
\label{sec:disc}

\subsection{Model uncertainties and limitations}

Whether or not our composite leaking H\,\textsc{ii} region and in situ HOLMES photoionization model can reproduce the enhanced $\eta$-parameter present in the eDIG of ESO~544-27 and PGC~28308 depends primarily on the $\Phi_\mathrm{HOLMES}/\Phi_\mathrm{OB}$ ratio. This ratio is sensitive to choices made during stellar population modeling as well as to SFR and $M_T$ measurements. To gain an understanding of the sensitivity of the \textsc{Starburst99} OB star population model to different parameters we tested the effect of different IMFs and evolutionary tracks to the predicted $\Phi_\mathrm{OB}$. We found that a model with steeper IMF for massive stars ($\alpha = 2.7$ rather than the standard Kroupa $\alpha = 2.3$), predicts $\Phi_\mathrm{OB}$ one third of our adopted model. As low SFR environment may give rise to a steeper IMF \citep[e.g.,][]{pflamm2008imf,rautio2024}, we may be overestimating $\Phi_\mathrm{OB}$ for ESO~544-27. While most of the alternative evolutionary tracks available in \textsc{Starburst99} did not produce large differences to the predicted $\Phi_\mathrm{OB}$, using the Geneva stellar tracks with rotation \citep{ekstrom2012tracks} caused a $\sim 70 \%$ increase in $\Phi_\mathrm{OB}$. The uncertainties in our SFR measurements are demonstrated by the differences we found between the measurements done using MUSE data and the measurements done using narrow-band data, the largest of which was $\sim 30 \%$ for IC~217. Additionally, H$\alpha$ alone as a SFR indicator tends to underpredict the true SFR for low luminosity galaxies \citep{lee2009hafuv}. The thick disk masses were obtained by fitting the 3.6 $\mu$m vertical surface-brightness profiles of each galaxy with the superposition of two isothermal disks in a hydrostatic equilibrium, and assuming a ratio of thick disk mass-to-light ratio to thin disk mass-to-light ratio of 1.2. This procedure gives a typical uncertainty of $\sim 50 \%$ to the resulting $M_\mathrm{T}$ \citep{comeron2012breaks}. Considering these uncertainties we determine that the parameters we used for the high-$\Phi_\mathrm{HOLMES}$ ESO~544-27 model are reasonable, while it is unlikely that $\Phi_\mathrm{HOLMES}/\Phi_\mathrm{OB}$ is high enough in PGC~28308 to explain its enhanced eDIG $\eta$-parameter.

Due to the one dimensional nature of our models, they fail to capture the nonlinear effects of the clumpy, turbulent 3D environment in which real eDIG clouds reside. This is evident from our inability to simultaneously reproduce the observed neutral [O\,\textsc{i}] line and the observed ionized lines ([N\,\textsc{ii}], [S\,\textsc{ii}], and [O\,\textsc{iii}]). {This limitation is imposed by the spatial and line-of-sight integration required in the Voronoi binning of our MUSE data that we use to fix many of the model parameters. Single Voronoi bin may contain, and its observed emission lines may originate from, ionized gas in many different conditions. Indeed, more detailed modeling of the small scale structure of the eDIG by post-processing density grids from magnetohydrodynamic (MHD) simulations has revealed the temporal and spatial variability of the density and ionization structure possible in eDIG and shown that turbulence and superbubbles can produce low-density channels through which ionizing photons can travel far above the midplane and photoionize eDIG \citep{barnes2014sim, barnes2015hds, vandenbroucke2018sim, kado-fon2020sim}. These models reproduce the vertically stratified multiphase ISM observed in the solar neighborhood with a few pc resolution, which is crucial in constructing an holistic model for the ionization of eDIG. However, none of these models based on MHD density grids have to our knowledge included HOLMES, and they struggle to reproduce the observed line ratios in eDIG. The next step in modeling eDIG would be the inclusion of ionizing radiation from HOLMES and rigorous photoionization modeling in MHD density grid post-processing.} Nevertheless, our 1D models are well suited to predicting the average integrated behavior of [N\,\textsc{ii}], [S\,\textsc{ii}], and [O\,\textsc{iii}] lines as a function of $z$, as the primary determinants of these lines, $\Phi_\mathrm{OB}$, $\Phi_\mathrm{HOLMES}$, and $n$, follow in real galaxies on average the same relations as they do in our models.

\subsection{Nature of eDIG ionization}

\begin{figure}
  \centering
  \includegraphics[width=0.5\textwidth]{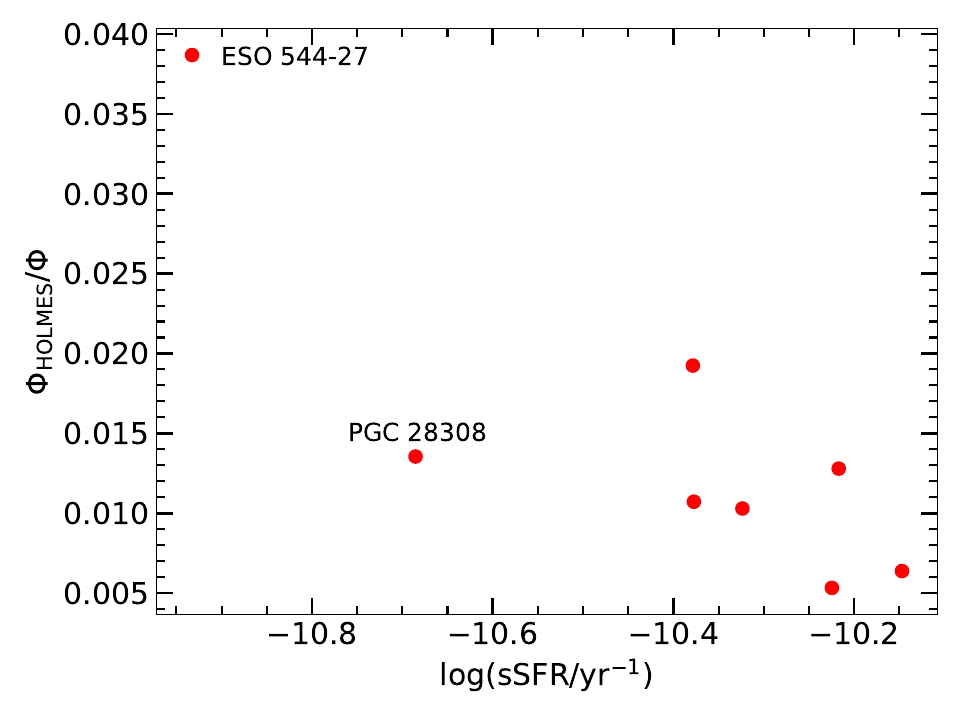}
  \caption{Relative contribution of HOLMES to the model ionizing radiation in eDIG against galaxy sSFR. ESO~544-27 and PGC~28308 are labeled. There is a clear anti-correlation between sSFR and $\Phi_\mathrm{HOLMES}/\Phi$ ($\rho = -0.79$, $p = 0.02$).}
  \label{fig:phisfr}
\end{figure}

Our results suggest that a combination of OB star radiation escaping the midplane H\,\textsc{ii} regions and in situ HOLMES radiation is capable of producing the enhanced high-ionization lines and $\eta$-parameter observed in the eDIG of galaxies such as ESO~544-27. However, this seems to require a HOLMES population that is sufficiently massive in relation to the OB star population, indicating a low-SFR galaxy. ESO~544-27 seems to be one such galaxy, as it is not part of the main sequence of star forming galaxies, but rather is in the green valley \mbox{\citepalias{rautio2022}}. We plot the relative contribution of $\Phi_\mathrm{HOLMES}$ to the ionizing radiation incident at the last cloud layer in the original models against the specific SFR (sSFR) of each galaxy in Fig. \ref{fig:phisfr}. We define the sSFR as sSFR$=$SFR/$(M_\mathrm{t}+M_\mathrm{T}+M_\mathrm{CMC})$, where $M_\mathrm{t}$ and $M_\mathrm{CMC}$ are the masses of the thin disk and central mass concentration, respectively, both from \mbox{\cite{comeron2018thick}}. Using the definition of \mbox{\cite{hsieh2017manga}} that star-forming galaxies have sSFR $>10^{-10.6}$ yr$^{-1}$, PGC~28308 also falls into the green valley. However, it also has the least massive thick disk in relation to its thin disk in our sample, and the high-ionization lines and the $\eta$-parameter are even more enhanced in the eDIG of PGC~28308 than the eDIG of ESO~544-27. Due to this, it seems unlikely that HOLMES and leaky H\,\textsc{ii} regions could alone explain the ionization state in the eDIG of PGC~28308.

\begin{figure*}
    \centering
    \includegraphics[width=\textwidth]{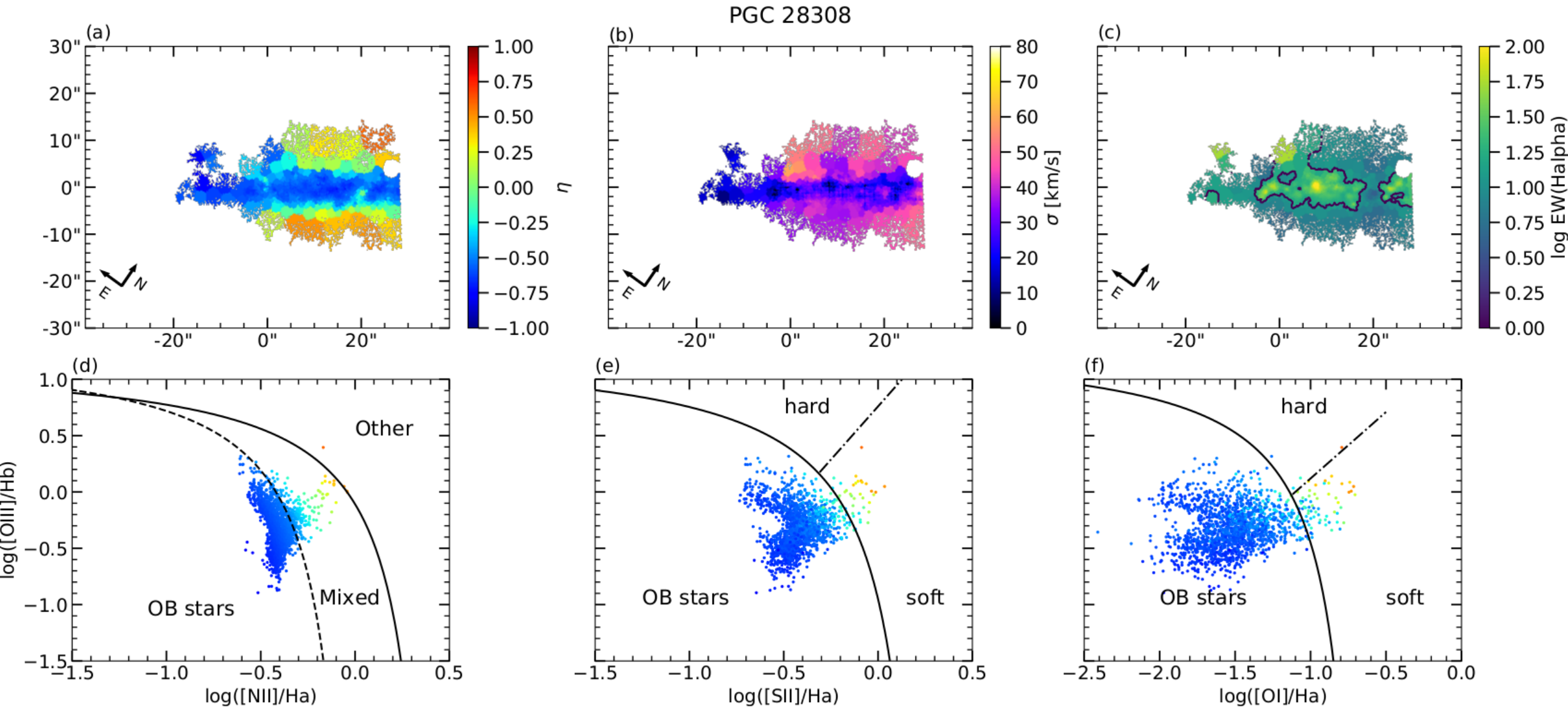}
    \caption{Emission-line properties for PGC~28308. \emph{(a)}: $\eta$-parameter map.  \emph{(b)}: velocity dispersion map. \emph{(c)}: logarithmic EW(H$\alpha$) map. The lower limit of the OB-star-dominated ionization (EW(H$\alpha$) = 14 Å) is contoured in black. \emph{(d), (e), (f)}: VO diagrams. The solid line is the \cite{kewley2001starburst} extreme starburst line, the dashed line in (d) is the \cite{kauffmann2003sfline} empirical starburst line, and the dash-dot lines in (e) and (f) are the \cite{kewley2006agn} Seyfert and low-ionization narrow emission-line region (LINER) demarcation line. The colors in the VO diagrams correspond to the colors in the $\eta$-parameter map. While there is no clear biconical morphology in the $\eta$-parameter map, the concentration of the eDIG over the center of the galaxy could hint at ionization by outflow driven shocks. Similar figures for ESO~443-21 and PGC~28308 are found in Appendix \ref{a:figs}.}
    \label{fig:lmap2}
\end{figure*}

Figure \ref{fig:phisfr} also demonstrates the anti-correlation between $\Phi_\mathrm{HOLMES}/\Phi$ in the eDIG and sSFR in our sample. We calculated the Spearman rank correlation coefficient for this relation $\rho = -0.79$, and obtained $p = 0.02$ as the $p$-value for the null hypothesis of no correlation. Due to the short lifetimes of OB stars ($\sim 10$ Myr), galaxies with low sSFR have relatively few OB stars active at any given moment compared to high-sSFR galaxies. HOLMES on the other hand have lifetimes larger than the age of the universe, and as such their numbers are independent of present day sSFR. Thus, it arises naturally that the photoionization of the eDIG by HOLMES is more significant in low-sSFR galaxies.

The case of eDIG ionization by a combination of in situ HOLMES radiation and leaking midplane H\,\textsc{ii} region radiation was first quantitatively explored by \mbox{\cite{flores-fajardo2011holmes}}, using a grid of photoionization models to reproduce the observed line ratios in the edge-on spiral galaxy NGC~891. In their models the propagation and transmission of radiation through the eDIG was not simulated, instead $\Phi_\mathrm{HOLMES}$ was fixed and $\Phi_\mathrm{OB}$ was varied, with the different values of $\Phi_\mathrm{OB}$ representing different amounts of absorption suffered by the escaping midplane OB star radiation. They concluded that when restricting to solar metallicity, models that fit the observations become dominated by HOLMES as $z$ increases. NGC~891 does have significantly enhanced [N\,\textsc{ii}], [S\,\textsc{ii}], and [O\,\textsc{iii}] in its eDIG similarly to ESO~544-27 and PGC~28308 \mbox{\citep{otte2001dig}}, but unlike them it has a relatively high SFR of 5.2 $M_\odot$ yr$^{-1}$ and low $M_\mathrm{T}$ of $3 \times 10^{10} M_\odot$ \mbox{\citep{flores-fajardo2011holmes}}. In fact it is likely that our model, if applied to NGC~891, would be able to reproduce the observed line ratios only when assuming a HOLMES population much more massive than that suggested by the $M_\mathrm{T}$ of NGC~891. This is because our model takes into account the dilution and absorption of HOLMES radiation unlike that of \mbox{\cite{flores-fajardo2011holmes}}, significantly reducing the incident $\Phi_\mathrm{HOLMES}$. As an example, for our high-$\Phi_\mathrm{HOLMES}$ ESO~544-27 model the sum of the incident $\Phi_\mathrm{HOLMES}$ and the transmitted incident $\Phi_\mathrm{HOLMES}$ for each cloud layer ranges from 2.5 to 25 times smaller than the total source $\Phi_\mathrm{HOLMES}$, which \mbox{\citep{flores-fajardo2011holmes}} used as the incident $\Phi_\mathrm{HOLMES}$ in their models. Therefore \mbox{\citep{flores-fajardo2011holmes}} may be overestimating $\Phi_\mathrm{HOLMES}$ by more than an order of magnitude.

If combination of in situ HOLMES and escaping midplane radiation is incapable of producing the ionization state in the eDIG of such galaxies as PGC~28308 and potentially NGC~891, an additional ionization source is required. Outflow driven shock ionization is one candidate. While PGC~28308 does not have clear biconical outflows like ESO~157-49 and IC~1553, the enhanced $\eta$-parameter and in fact the eDIG altogether as detected by MUSE, is concentrated above the center of PGC~28308. Moreover, as the MUSE data does not cover the full radial extent of the galaxy, we do not have the whole picture of the eDIG morphology. The ionization state in PGC~28308 eDIG could be caused by an advanced stage of a large superbubble breakout, or combination of multiple smaller superbubbles breakouts. Figure \ref{fig:lmap2} shows the $\eta$-parameter map, ionized gas velocity dispersion ($\sigma$) map, the H$\alpha$ equivalent width map, and the Veilleux-Osterbrock (VO; \citealt{veilleux1987vo}) suite of diagnostic diagrams for PGC~28308. The slight enhancement of velocity dispersion in PGC~28308 eDIG could be caused by shocks. The low H$\alpha$ equivalent width (EW(H$\alpha) < 14$ Å) in PGC~28308 eDIG indicates that OB star radiation alone cannot be responsible for its ionization. Similar situation may contribute to the enhanced line ratios in NGC~891 eDIG, as evidence of a superbubble has been found therein \mbox{\citep{yoon2021bubble}}.

Our small sample of eight edge-on galaxies gives likely examples of at least three distinct scenarios of eDIG ionization:
\begin{enumerate}
\item For half of the galaxies (ESO~443-21, ESO~469-15, IC~217, and PGC~30591) their eDIG emission can be satisfactorily explained with just photoionization by leaking radiation from midplane H\,\textsc{ii} regions.
  \item To explain the ionization state of the eDIG of the green valley galaxy ESO~544-27, photoionization by in situ HOLMES is required in addition to photoionization by leaking H\,\textsc{ii} regions.
\item In ESO~157-49, IC~1553, and potentially PGC~28308, superbubble driven shock ionization causes harder ionization and elevates the $\eta$-parameter in certain parts of the eDIG.
\end{enumerate}
All three scenarios are also supported by other recent work. \mbox{\cite{levy2019edge}} finds photoionization by leaking H\,\textsc{ii} regions to account for $90\%$ of the eDIG ionization in their sample of 25 edge-on galaxies selected from the Calar Alto Legacy Integral Field Area (CALIFA) survey, while \mbox{\cite{lacerda2018ew}} finds four edge-on galaxies where above and below the plane all emission indicated photoionization dominated by HOLMES also from CALIFA data. Both \mbox{\cite{levy2019edge}} and \mbox{\cite{lacerda2018ew}} use low H$\alpha$ equivalent width as an indicator of HOLMES ionization, indicating that the differences between their results come indeed from differences between their samples rather than differences between their methodologies. Also from CALIFA \mbox{\cite{lopez-coba2019outflows}} finds evidence of outflows in 17 edge-on galaxies. Similar composite ionization models with leaking H\,\textsc{ii} regions and HOLMES as ionization sources are found to also work well for DIG in face-on galaxies \mbox{\citep{belfiore2022dig}}.

\section{Summary and conclusions}
\label{sec:sum}
We constructed a self-consistent\footnote{in plane-parallel approximation} model for the photoionization of eDIG by leaking midplane H\,\textsc{ii} regions and in situ HOLMES using the \textsc{Cloudy} photoionization code. Our model rigorously simulates the dilution and procession of the radiation of both HOLMES and midplane OB stars as it propagates in the eDIG. We fit the model to the observed vertical profiles of [N\,\textsc{ii}]/H$\alpha$, [S\,\textsc{ii}]/H$\alpha$, and [O\,\textsc{iii}]/H$\beta$ line ratios of the eight edge-on galaxies in our MUSE sample.

We find that leaking radiation of midplane H\,\textsc{ii} regions is sufficient to explain the line ratios in the eDIG where $\eta < -0.5$, which is most of the eDIG in ESO~157-49, ESO~443-21, ESO~469-15, IC~217, IC~1553, and PGC~30591, beside the biconical shocked regions in ESO~157-49 and IC~1553. For the green valley galaxy ESO~544-27, we constructed a high-$\Phi_\mathrm{HOLMES}$ model with maximized $\Phi_\mathrm{HOLMES}$ and minimized $\Phi_\mathrm{OB}$ that were still within the uncertainties of our observations and modeling choices. This model was able to reproduce the $\eta > -0.5$ observed in the eDIG of ESO~544-27. {We confirmed that inclusion of HOLMES improves the fit by $\chi^2$ minimization.} We constructed a similar model for PGC~28308, but found that it required unreasonably high $\Phi_\mathrm{HOLMES}$/$\Phi_\mathrm{OB}$ to reproduce the observed line ratios and $\eta$-parameter. We speculate that outflow driven shock ionization could be responsible for the ionization state of PGC~28308 instead. We also show that there exists a clear anti-correlation between sSFR and $\Phi_\mathrm{HOLMES}$/$\Phi$ for our sample galaxies. 

Our work here and in \citetalias{rautio2022}, as well as other recent studies \citep{lacerda2018ew, levy2019edge, lopez-coba2019outflows}, support a great variance in the ionization sources of eDIG. Leaking midplane radiation, in situ HOLMES, and shocks all contribute to eDIG ionization to different degrees in different galaxies. Our work seems to suggest that the contribution of HOLMES negatively correlates with the sSFR of a galaxy, with the low-SFR galaxy ESO~544-27 being the only one with significant HOLMES ionization. Further work is required with larger samples to confirm this anti-correlation and search for other correlations between the ionization sources and galaxy properties.

\begin{acknowledgements}
  {We thank the referee on useful comments that helped us perform more rigorous fitting.}
  This research is based on observations collected at the European Southern Observatory under ESO programmes 096.B-0054(A) and 097.B-0041(A).
  RR acknowledges funding from the Technology and Natural Sciences Doctoral Program (TNS-DP) of the University of Oulu, and from the Vilho, Yrjö and Kalle Väisälä Foundation of the Finnish Academy of Science and Letters.
  AW acknowledges support from the STFC [grant numbers ST/S00615X/1 and ST/X001318/1].
  SC acknowledges funding from the State Research Agency (AEI-MCINN) of the Spanish Ministry of Science and Innovation under the grant  “Thick discs, relics of the infancy of galaxies" with reference PID2020-113213GA-I00.
  AV acknowledges funding from the Academy of Finland grant n:o 347089.
 This research made use of python (\url{http://www.python.org}), SciPy \citep{2020SciPy}, NumPy \citep{2020NumPy}, Matplotlib \citep{matplotlib}, and Astropy, a community-developed core Python package for Astronomy. 
\end{acknowledgements}

\bibliographystyle{aa}
\bibliography{ref}{}

\begin{appendix}
  
  \section{Effects of different parameters on the model}
  \label{a:vary}

\begin{figure}[h]
    \centering
    \includegraphics[width=0.5\textwidth]{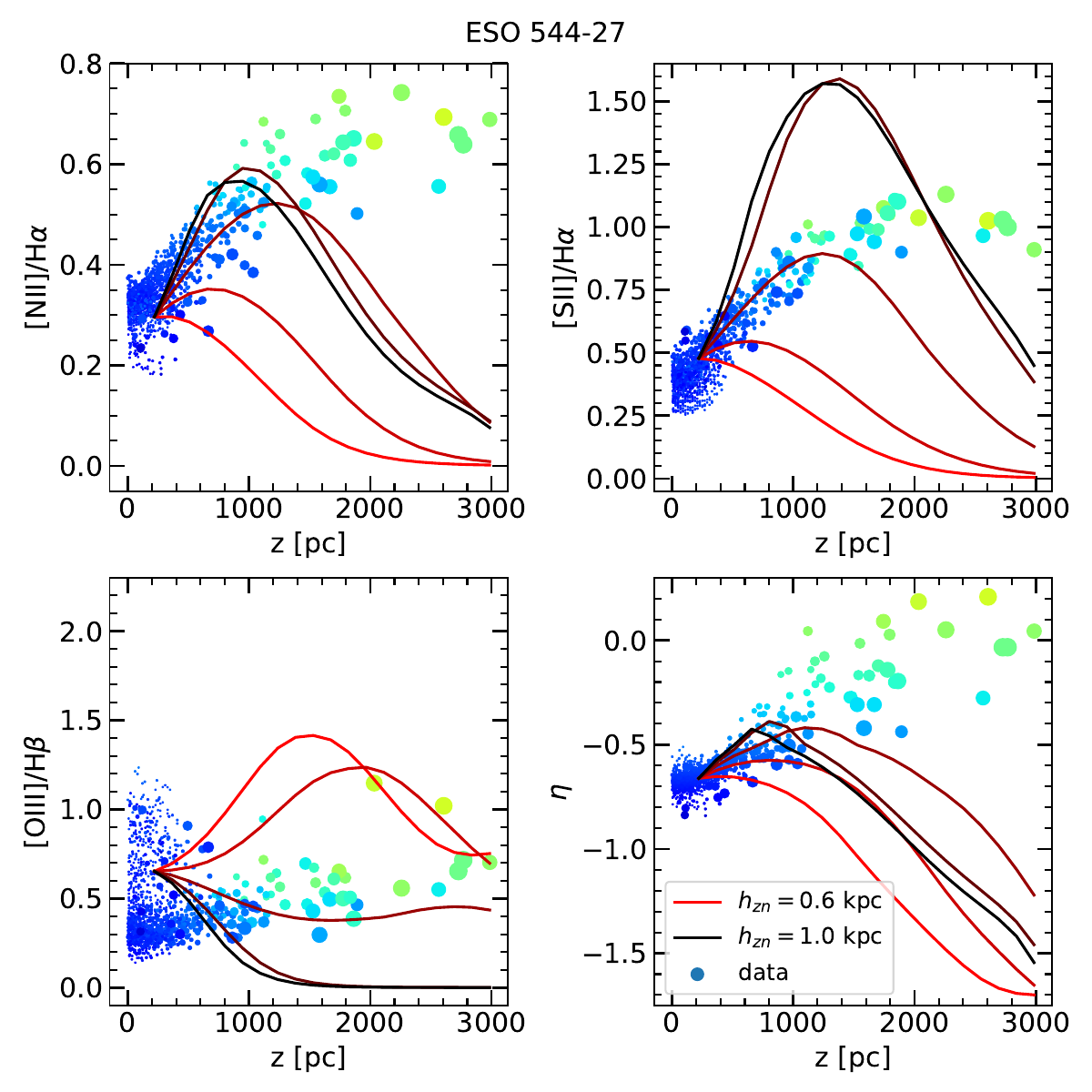}
    \caption{Line ratio and $\eta$-parameter profiles for ESO~544-27 models with varying $h_{zn}$. The dots show the observed values, colored according to the $\eta$-parameter going from blue at low $\eta$ to green at high $\eta$. The size of each dot is relative to the size of the corresponding Voronoi bin. The curves show the model predictions with the color of the curve corresponding with the model $h_{zn}$, ranging from $h_{zn} = 0.6$ kpc. (red) to $h_{zn} = 1.0$ kpc (black) with $h_{zn} = 0.1$ kpc steps.}
    \label{fig:varyz0}
  \end{figure}

\begin{figure}
    \centering
    \includegraphics[width=0.5\textwidth]{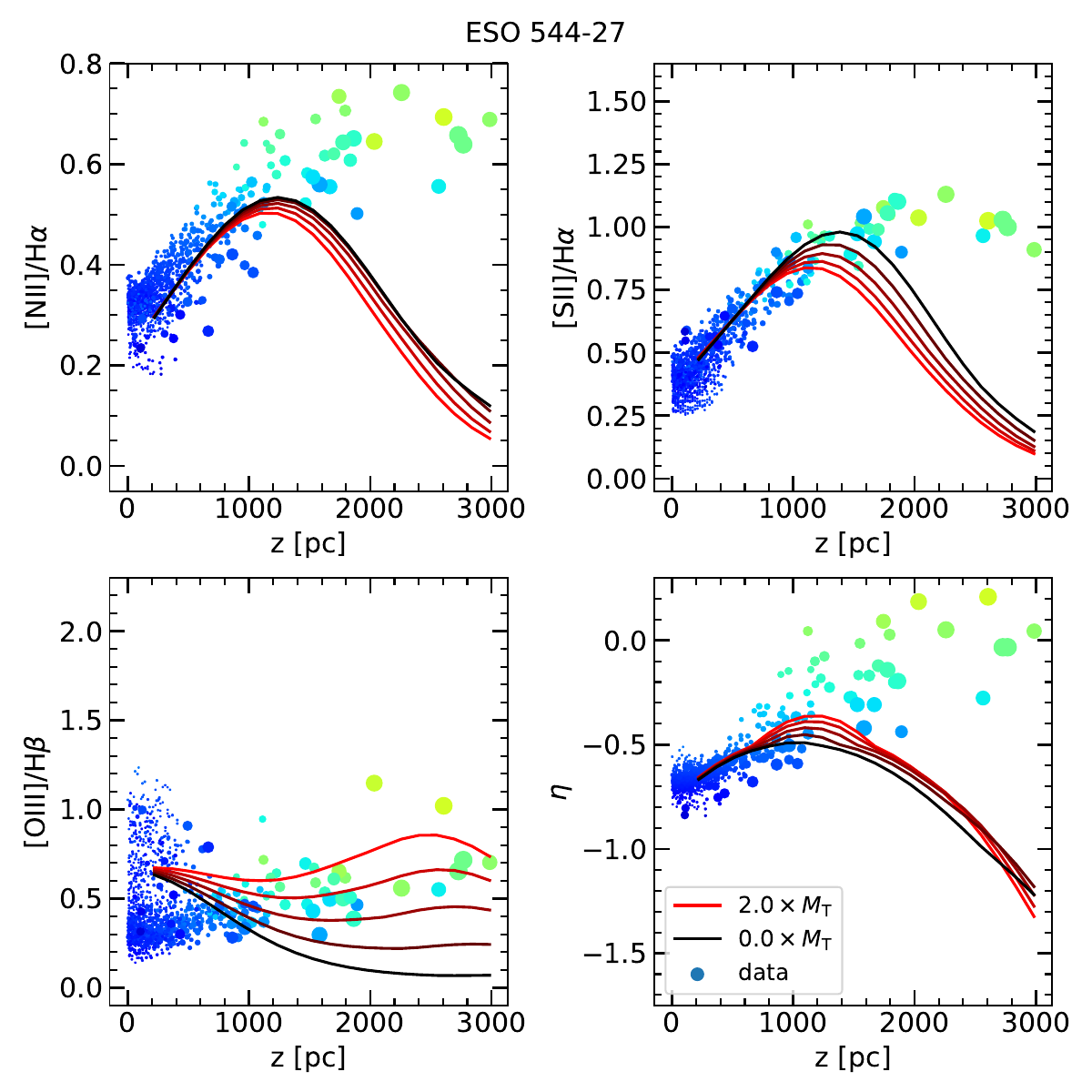}
    \caption{Line ratio and $\eta$-parameter profiles for ESO~544-27 models with varying HOLMES mass. The dots show the observed values, colored according to the $\eta$-parameter going from blue at low $\eta$ to green at high $\eta$. The size of each dot is relative to the size of the corresponding Voronoi bin. The curves show the model predictions with the color of the curve corresponding with the model $M_{\mathrm{T}}$, ranging from $0.0 \times M_{\mathrm{T}}$. (black) to $2.0 \times M_{\mathrm{T}}$ (red) with $0.5 M_{\mathrm{T}}$ steps.}
    \label{fig:varyH}
  \end{figure}

\begin{figure}
    \centering
    \includegraphics[width=0.5\textwidth]{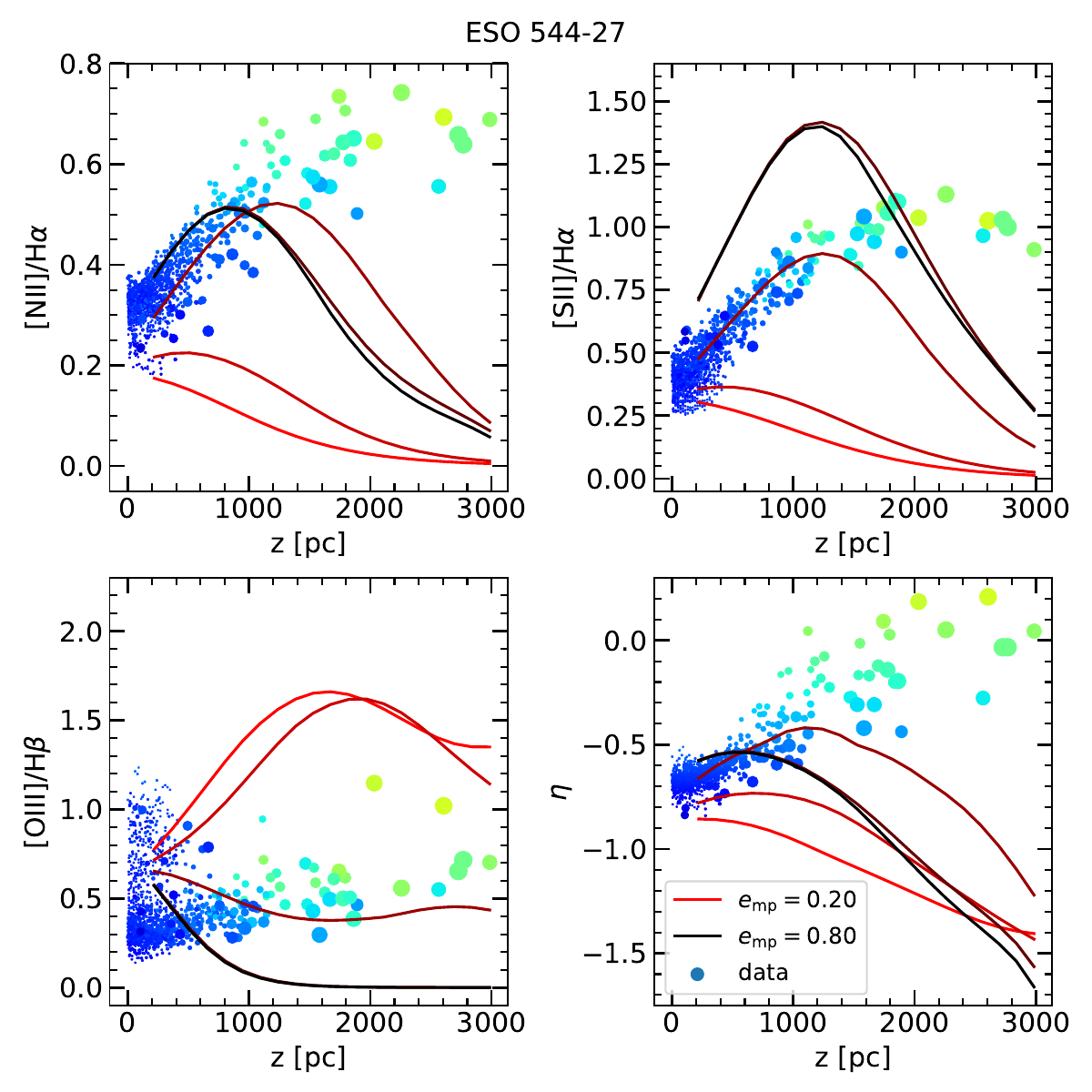}
    \caption{Line ratio and $\eta$-parameter profiles for ESO~544-27 models with varying midplane escape fraction. The dots show the observed values, colored according to the $\eta$-parameter going from blue at low $\eta$ to green at high $\eta$. The size of each dot is relative to the size of the corresponding Voronoi bin. The curves show the model predictions with the color of the curve corresponding with the model $e_\mathrm{mp}$, ranging from $e_\mathrm{mp} = 0.2$. (red) to $e_\mathrm{mp} = 0.8$ (black) with $e_\mathrm{mp} = 0.15$ steps.}
    \label{fig:varyef}
  \end{figure}

\begin{figure}
    \centering
    \includegraphics[width=0.5\textwidth]{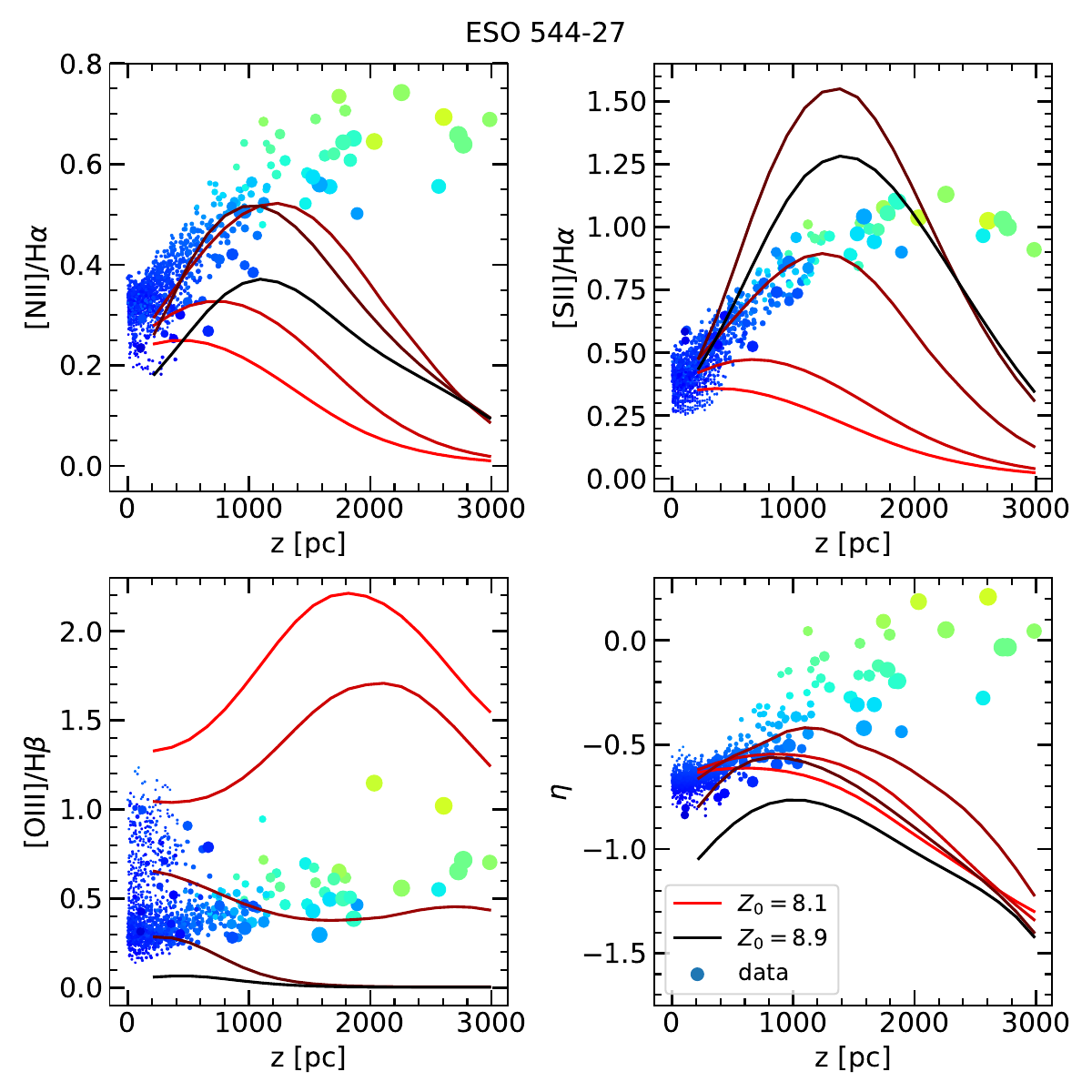}
    \caption{Line ratio and $\eta$-parameter profiles for ESO~544-27 models with varying metallicity. The dots show the observed values, colored according to the $\eta$-parameter going from blue at low $\eta$ to green at high $\eta$. The size of each dot is relative to the size of the corresponding Voronoi bin. The curves show the model predictions with the color of the curve corresponding with the model $Z_0$, ranging from $Z_0 = 8.1$. (red) to $Z_0 = 8.9$ (black) with $Z_0 = 0.2$ steps.}
    \label{fig:varyZ}
  \end{figure}

\begin{figure}
    \centering
    \includegraphics[width=0.5\textwidth]{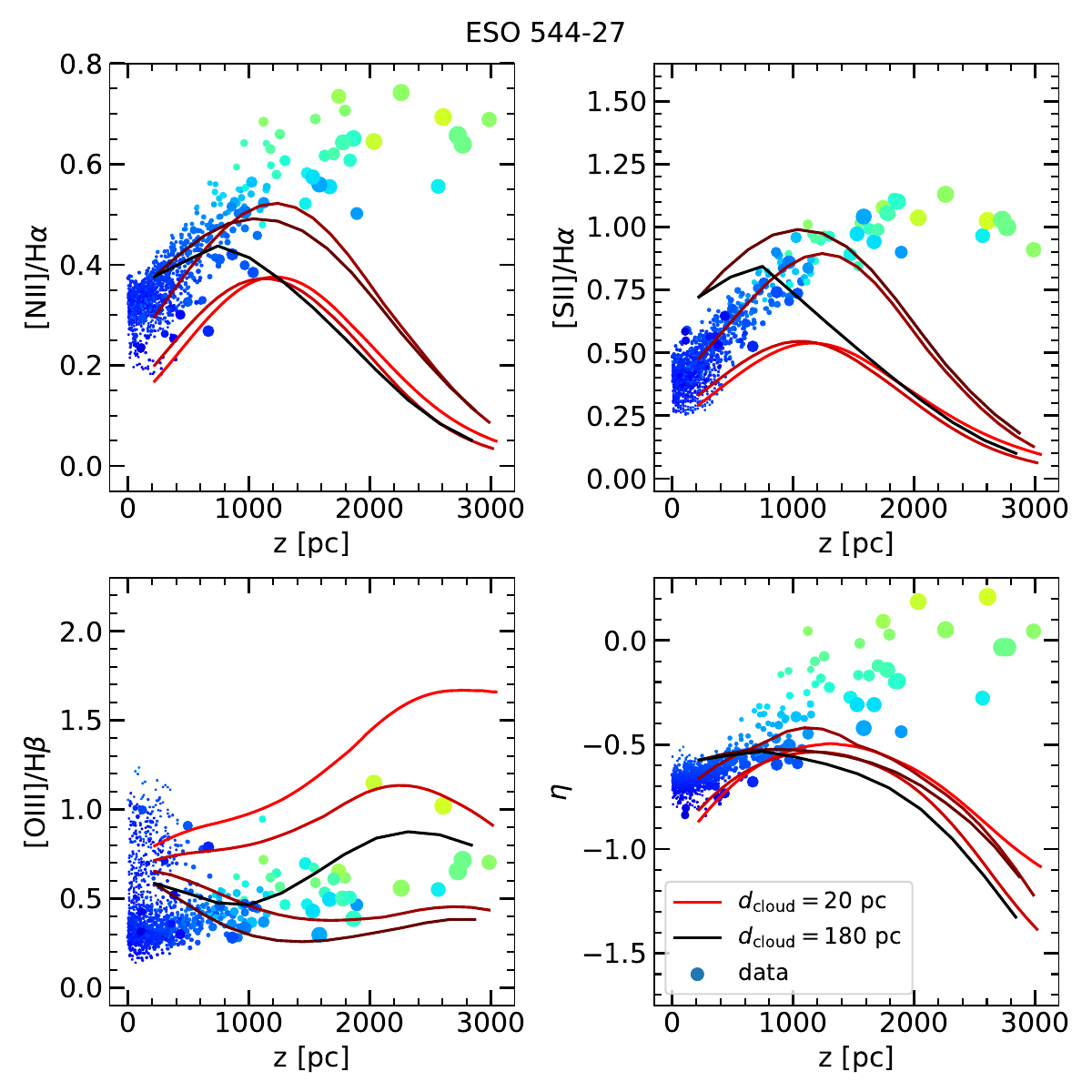}
    \caption{Line ratio and $\eta$-parameter profiles for ESO~544-27 models with varying cloud size. The dots show the observed values, colored according to the $\eta$-parameter going from blue at low $\eta$ to green at high $\eta$. The size of each dot is relative to the size of the corresponding Voronoi bin. The curves show the model predictions with the color of the curve corresponding with the model $d_\mathrm{cloud}$, ranging from $d_\mathrm{cloud} = 20$. pc (red) to $d_\mathrm{cloud} = 180$ pc (black) with $d_\mathrm{cloud} = 40$ pc steps.}
    \label{fig:varycs}
  \end{figure}

  We present here figures similar to Figs. \ref{fig:varyU} and \ref{fig:varyff} for other parameters of our model. All of the models presented in the figures use ESO~544-27 data as a baseline and have all but one parameter fixed. Figure \ref{fig:varyz0} shows models with varying $h_{zn}$, Fig. \ref{fig:varyH} shows models with varying HOLMES mass, Fig. \ref{fig:varyef} shows models with varying midplane escape fraction, Fig. \ref{fig:varyZ} shows models with varying metallicity, and Fig. \ref{fig:varycs} shows models with varying eDIG cloud size. The fixed parameters are $\log U_1 = -3.01$, $f_V = 0.17$, $h_{zn} = 0.8$ kpc, $e_\mathrm{mp} = 0.5$, $Z_0 = 8.5$, and $d_\mathrm{cloud} = 100$ pc. It is clear from the figures that slight variations in one parameter can easily be compensated by other parameters. Therefore, while there are significant uncertainties in many of the parameters that we derive from data, our model is robust.  

  \FloatBarrier

  \clearpage
  
  \section{Images of ESO~443-21, PGC~28308, and PGC~30591}
  \label{a:figs}

\begin{figure*}[b]
    \centering
    \includegraphics[width=\textwidth]{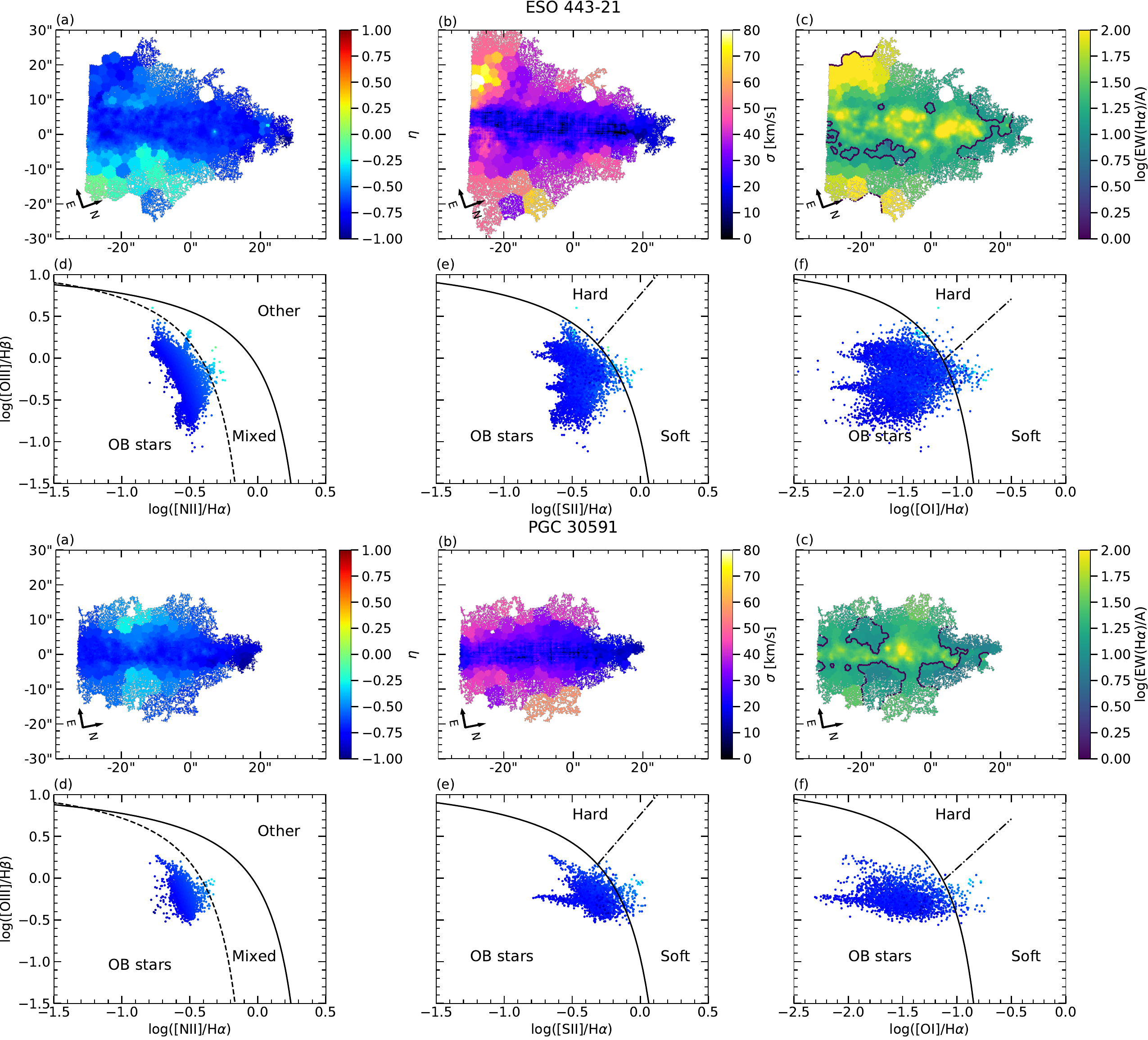}
    \caption{Same as Fig. \ref{fig:lmap2} but for ESO~443-21 and PGC~30591.}
    \label{fig:lmap0}
  \end{figure*}

\begin{figure}
    \centering
    \includegraphics[width=0.5\textwidth]{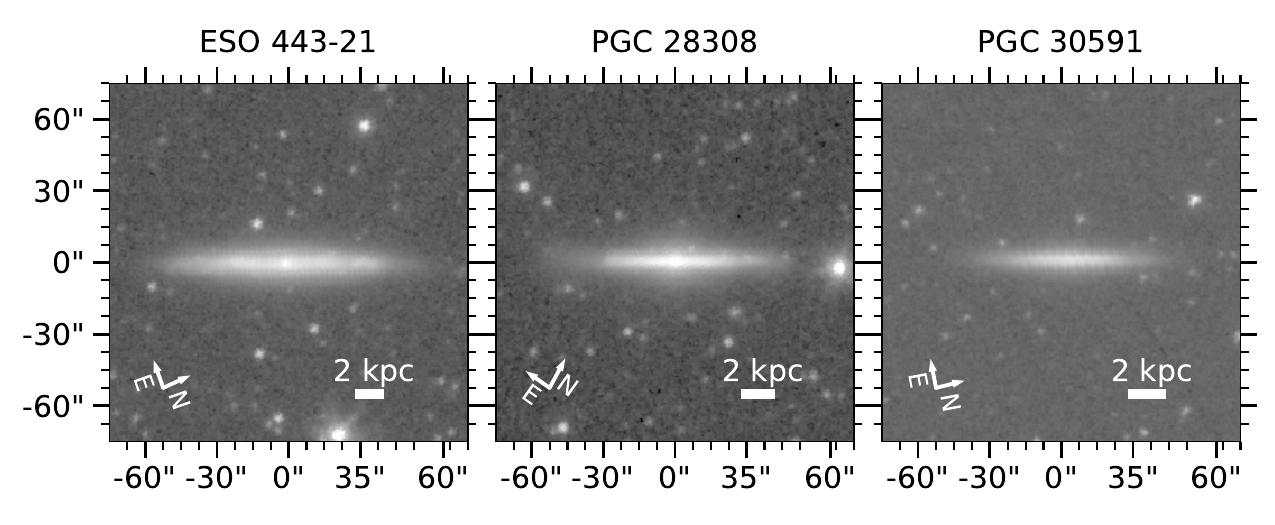}
    \caption{S$^4$G \citep{sheth2010s4g} 3.6 micron images of ESO~443-21, PGC~28308, and PGC~30591. Directions on the sky and 2 kpc scale bars are shown in the images.}
    \label{fig:s4g}
  \end{figure}

  For completeness, we present here some images of the three galaxies that were not part of the sample used in \citetalias{rautio2022} (ESO~443-21, PGC~28308, and PGC~30591). Figure \ref{fig:lmap0} shows the $\eta$-parameter maps, the H$\alpha$ equivalent width maps, ionized gas velocity dispersion ($\sigma$) maps, and the Veilleux-Osterbrock (VO; \citealt{veilleux1987vo}) suite of diagnostic diagrams for ESO~443-21 and PGC~30591. For PGC~28308 these were already presented in Fig.~\ref{fig:lmap2}. Figures~\ref{fig:lmap2} and \ref{fig:lmap0} were derived from the MUSE cubes in the same way as Fig.~14 in \citetalias{rautio2022}. The S$^4$G 3.6 micron images of ESO~443-21, PGC~28308, and PGC~30591 are shown in Fig.~\ref{fig:s4g}.
  
\end{appendix}

\end{document}